\documentclass[english,a4paper,11pt]{article}
\usepackage[margin=3cm]{geometry}
\usepackage[latin1]{inputenc}
\usepackage{latexsym}
\usepackage{booktabs}
\usepackage{amsmath}
\usepackage{eurosym}
\usepackage{textcomp}
\usepackage{algorithm}
\usepackage{makecell}
\usepackage{latexsym}
\usepackage{pdflscape}
\usepackage[final]{graphicx}
\DeclareGraphicsExtensions{.jpg,.jpeg,.pdf,.png,.mps}
\usepackage{epsfig}
\usepackage[round]{natbib}
\usepackage{rotating}
\usepackage{color}
\usepackage{graphicx,subfig}
\usepackage{colortbl}
\usepackage{float}

\usepackage{csquotes}
\usepackage[misc]{ifsym}
\usepackage{multirow}
\usepackage[toc,page]{appendix}
 \usepackage{threeparttable}
\usepackage[bitstream-charter]{mathdesign}
\usepackage[T1]{fontenc}
\usepackage{lmodern}
\usepackage{tikz}
\usepackage{textcomp}
\usetikzlibrary{arrows.meta,positioning,shapes.geometric,calc}
\usepackage{xcolor}
\usepackage[
    colorlinks=true,
    citecolor=blue,
    linkcolor=blue,
    urlcolor=blue
]{hyperref}

\title{\Large Hybrid Time Series Models for Short-Term Sea-Level Forecasting}

\author{$\mathrm{Pierdomenico \ Duttilo}^\mathrm{*,\hspace{0.5mm}\textrm{\Letter}},  
	\  \mathrm{Francesco \ Lisi}^\mathrm{*}$
	\\  
	$^\mathrm{*}$\small{\emph{Department  of Statistical Sciences,  University of Padua, Italy}}\\
    $^\mathrm{\textrm{\Letter}}$\small{\emph{Corresponding author: \href{mailto:pierdomenico.duttilo@unipd.it}{\textcolor{blue}{pierdomenico.duttilo@unipd.it}}}}
}
\date{}

\begin{document}
	\maketitle

\begin{abstract}
Accurate tide forecasts are essential for coastal management, navigation, flood-risk reduction, and infrastructure protection. Observed sea level can be decomposed into astronomical and non-astronomical components, the latter mainly driven by meteorological effects. This study investigates a hybrid framework for hourly sea-level forecasting that combines harmonic analysis (HA) for the astronomical component with data-driven models for the non-astronomical contribution. The approach is evaluated at six tide-gauge stations with different tidal regimes: Venice, Trieste, Saint-Malo, Vard{\o}, Nikiski, and Nagasaki. Four data-driven model classes are considered: (i) autoregressive and (ii) functional autoregressive models with exogenous variables; (iii) generalized additive models and autoregressive neural networks; and (iv) a novel semi-functional k-nearest-neighbours approach that combines similarity in recent non-astronomical trajectories and meteorological conditions. Point forecasts are complemented by upper quantile forecasts, obtained by historical simulation of out-of-sample errors, targeting the right tail of the sea-level distribution. At the 24-hour horizon, hybrid models reduce the mean absolute and root mean squared errors of HA by 54.9\% and 52.9\% on average, while the best hybrid model reduces the HA quantile score by 40--60\%. HA systematically underestimates the right tail. No hybrid specification dominates across locations, although the generalized additive model is the most competitive overall. For Venice, replacing observed meteorological covariates with operational Global Forecast System weather forecasts leaves 24- and 48-hour accuracy virtually unchanged. A cost--loss analysis of the MoSE flood-barrier system, based on these real-time forecasts, shows that hybrid forecasts substantially reduce decision losses relative to HA, and that activation rules based on upper quantile forecasts reduce them further for all models.
\noindent 
\vspace{0.50cm}
\vspace{0.40cm}
\end{abstract}

\section{Introduction}
Tides are periodic variations in sea level primarily driven by the gravitational attraction of the Moon and the Sun. Observed sea levels also reflect non-astronomical processes associated with atmospheric forcing, local bathymetry, coastal geometry, and water depth \citep{doodson1954,devlin2016,abubakar2019}. Their interaction generates different dynamics, making accurate forecasting important for navigation, port operations, coastal protection, and marine resource management \citep{hallegatte2011,makris2021,khojasteh2022}. Sea-level forecasting has traditionally relied on harmonic analysis (HA), which represents the regular tidal signal as the superposition of sinusoidal constituents. However, HA may be insufficient when meteorological forcing and local hydrodynamic processes contribute substantially to observed sea-level \citep{abubakar2019,umgiesser2021}. This has encouraged the use of data-driven models able to exploit temporal dependence, meteorological information, and nonlinear relationships \citep{ayinde2024}. More recently, hybrid approaches have combined HA for the astronomical component with data-driven models for the non-astronomical component \citep{ishida2020,jiao2026}.

This study investigates a hybrid framework for hourly sea-level forecasting at six tide-gauge stations located in Venice and Trieste (Italy), Saint-Malo (France), Vard{\o} (Norway), Nikiski (Alaska, USA), and Nagasaki (Japan). Observed sea level is decomposed into astronomical and non-astronomical components. The former is driven by gravitational forcing and estimated using HA, whereas the latter captures meteorological effects and other sources of local variability. The non-astronomical component is estimated by considering four main data-driven model classes: (i) linear parametric models, represented by autoregressive models with exogenous variables (ARX); (ii) functional parametric models, based on functional autoregressive models with exogenous variables (FARX); (iii) semi-parametric and nonlinear models, including generalized additive models (GAM) and autoregressive neural networks with exogenous inputs (NNARX); and (iv) a semi-functional model based on a non-standard k-nearest neighbours (KNN) approach. The KNN model combines similarity in recent non-astronomical trajectories and meteorological conditions. Meteorological information is included by wind and atmospheric pressure. The total sea-level forecast is obtained by combining the two component forecasts. Particular attention is devoted to Venice, a widely studied case in the literature, where high-water events have important consequences for economic activities \citep{petaccia2006,massalin2007,dinunno2021,umgiesser2021, Faranda2023,giupponi2024}. 

This work contributes to the literature in several ways. First, it assesses the predictive contributions of the astronomical and non-astronomical components of sea level across locations with different tidal regimes. This analysis identifies when HA alone provides adequate forecasts and when modelling meteorological forcing and residual dynamics yields additional predictive gains. Second, it compares the predictive accuracy of several hybrid models across multi-step forecast horizons. As part of this comparison, a tailored non-standard $k$-nearest-neighbours specification is introduced, combining similarities in recent non-astronomical trajectories and meteorological conditions. Third, point forecasts are complemented by upper-quantile forecasts generated by applying historical simulation to post-processed residuals. Their calibration and accuracy in the right tail of the sea-level distribution are assessed using exceedance frequencies and quantile scores. Fourth, for Venice, the forecast evaluation is extended to an economic decision-making framework for the MoSE (MOdulo Sperimentale Elettromeccanico, Experimental Electromechanical Module) flood-barrier system. Activation rules based on both point and upper quantile forecasts, obtained with operational weather forecasts, are compared within a cost--loss framework, thereby highlighting the economic value of accurate and probabilistic sea-level forecasts.

The remainder of the manuscript is organized as follows.  Section~\ref{sec:literature} reviews the main approaches to sea-level forecasting. Section~\ref{sec:data} describes the sea-level, meteorological, and site-specific data for the six tide-gauge stations. Section~\ref{sec:method} introduces the hybrid forecasting framework and the competing models for the non-astronomical component.  Section~\ref{sec:forecasting_design} describes the design of the forecasting experiment and the procedures used to evaluate predictive accuracy. Section~\ref{sec:forecasting_results} presents and compares the out-of-sample forecasting results across locations and forecast horizons. Section~\ref{sec:mose_cost_loss} examines the operational and economic implications of tide forecasting for decision-making in Venice. Section~\ref{sec:conclusions} summarizes the main findings and provides concluding remarks.

\section{Literature review}\label{sec:literature}
The literature on sea-level forecasting is extensive and encompasses a wide range of methods. It can be broadly classified into five categories: harmonic, autoregressive, hydrodynamic, machine/deep-learning, and hybrid approaches \citep{abubakar2019,ayinde2024}. Harmonic analysis (HA) is the earliest approach to tidal analysis and forecasting. Building on Newton's theory of gravitation and later developments by P.S. Laplace, W. Thomson, and G. Darwin, it represents tides as the superposition of sinusoidal constituents, each characterized by frequency, amplitude, and phase \citep{darwin1898tides,darwin1907scientific,abubakar2019,cai2018}. \cite{Doodson1921,doodson1954} further developed this framework by identifying 388 tidal constituents and introducing the Doodson-number classification. \cite{franco1971fft,franco1997tides,franco2009mares} developed frequency-domain methods for HA. Modern implementations can include hundreds of harmonic constituents, although only a limited subset is generally required in practice \citep{Marone2013,yin2015hybrid,stephenson2016,abubakar2019,jiao2026}. Extensions of the classical framework have also been developed to account for time-varying effects and nonstationary tidal regimes \citep{foreman2009versatile,matte2013adaptation,guo2018harmonic}. HA is physically interpretable and effective for astronomical tides. Its performance decreases when the non-astronomical component dominates or when records are too short to resolve closely spaced constituents \citep{abubakar2019,umgiesser2021}. Autoregressive-based models exploit the temporal dependence of observed sea-level to construct short-term forecasts \citep{tomasin1979,petaccia2006,massalin2007,chen2020autoregressive}. Early applications combined lagged sea-level observations with external meteorological variables \citep{tomasin1979,petaccia2006,massalin2007}. Autoregressive models are also used to correct temporally correlated residuals from astronomical tidal models \citep{chen2020autoregressive}.

Hydrodynamic models simulate water motion under astronomical, atmospheric, and oceanographic forcing \citep{massalin2007}. They represent processes such as wind stress, atmospheric pressure, circulation, bathymetry, and coastline geometry \citep{umgiesser2021}. In the northern Adriatic, examples include SHYFEM (\textit{Shallow water HYdrodynamic Finite Element Model}) and HYPSE (\textit{HYdrostatic Padua Sea Elevation}) model \citep{umgiesser1986model,umgiesser1993staggered,lionello1998surface,umgiesser2004finite,lionello2006data}. Similar models are used elsewhere: SCHISM (\textit{Semi-implicit Cross-scale Hydroscience Integrated System Model}) along the Portuguese Atlantic coast, ROMS (\textit{Regional Ocean Modeling System}) in Norway, and HAMSOM (\textit{Hamburg Shelf Ocean Model}) in Spanish coasts \citep{alvarez2001nivmar,fortunato2017,saetra2018}. Their main advantage is the physically based representation of sea-level dynamics. High-resolution simulations are computationally demanding \citep{umgiesser2021}.

The increasing availability of high-frequency data together with greater computational resources, has increased the use of machine-learning and deep-learning methods. Early studies mainly used artificial neural networks (ANNs) to capture nonlinear relationships in tidal and coastal water-level time series \citep{deo1998tide,tsai1999back}. Other approaches include support vector machines, neuro-fuzzy models, extreme learning machines, and relevance vector machines \citep{abubakar2019,ayinde2024}. Recent research has focused on recurrent and deep neural networks for longer temporal dependencies. \cite{dinunno2021} used NNARX with lagged sea-level and meteorological inputs for forecasts up to 72 hours ahead in the Venice lagoon. \cite{ishida2020} developed a long short-term memory (LSTM) model combining astronomical information with wind, atmospheric pressure, temperature, and climatic variables. LSTM-based models have also been applied at several Mediterranean locations \citep{accarino2021multimodel}. More complex architectures combine different neural components, such as temporal convolutional networks and LSTM \citep{almaliki2025}. These methods are flexible in representing nonlinear and nonstationary relationships but require attention to generalization and interpretability \citep{ayinde2024}.

Hybrid approaches combine the strengths of physical, harmonic, autoregressive, and machine/deep-learning methods. A common strategy is to model the astronomical component through HA and use a data-driven model for the residual component. Examples include HA combined with neural networks \citep{yin2015hybrid}, wavelets \citep{ElDiasty2018}, and statistical models \citep{abubakar2019}. Other approaches introduce physical or astronomical information directly into machine-learning predictors \citep{ishida2020}. More recently, physics-guided methods have incorporated physical knowledge into the learning process. \cite{jiao2026} included nonstationary HA in the loss function of a bidirectional gated recurrent unit network to improve physical consistency while retaining the flexibility of deep learning. Recently, \cite{DuttiloLisi2026} examined the contribution of meteorological variables to sea-level forecasting, focusing only on Venice, point forecasts, and neural network models. Within this restricted setting, they found that including meteorological covariates improved forecast accuracy. The present work extends the analysis by considering multiple tide-gauge stations with different tidal and meteorological regimes, probabilistic forecasting, and a broader set of model classes. In addition, the practical relevance is illustrated through an economic evaluation of the forecasts for the operation of the MoSE barrier in Venice.

\section{Data}\label{sec:data}
\subsection{Sea-level data}
\begin{figure}[H]
    \centering
    \includegraphics[width=0.85\textwidth]{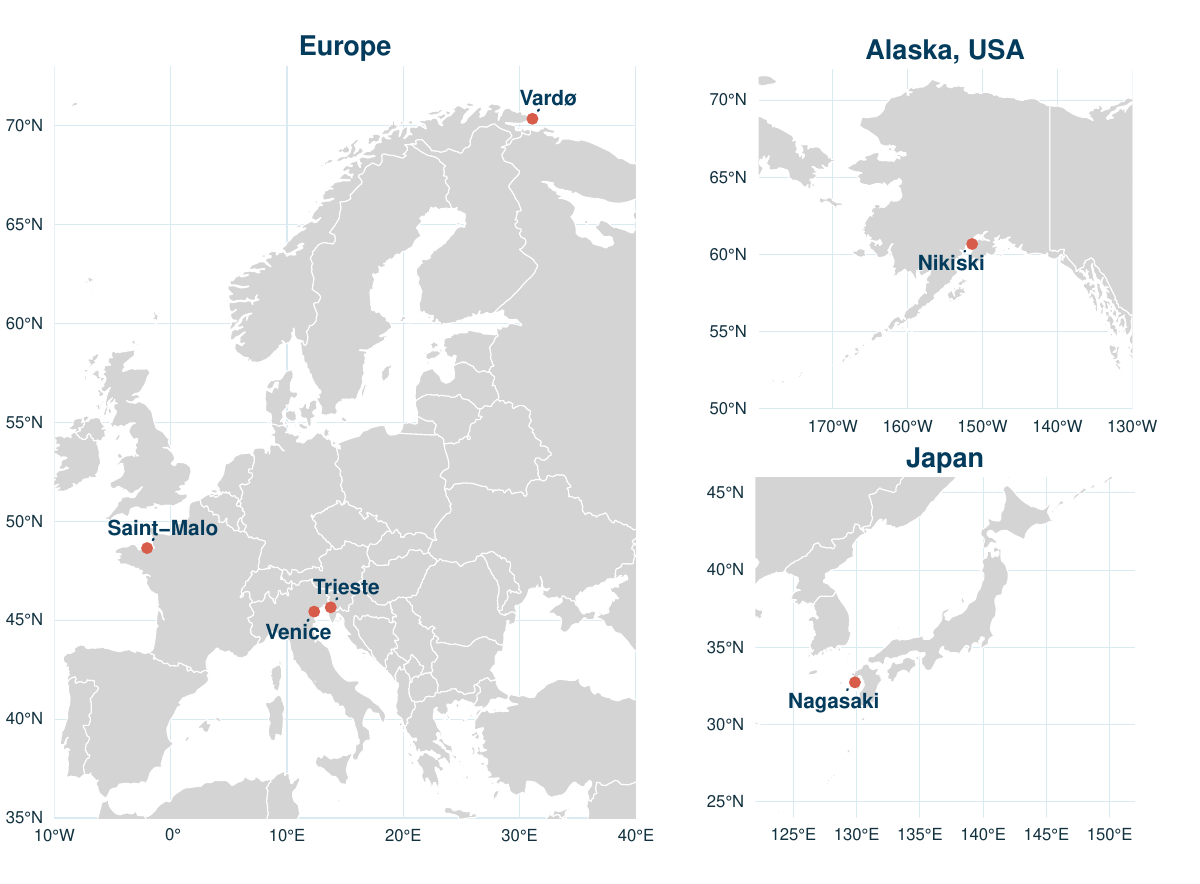}
    \caption{Geographical location of the six tide-gauge stations (red dots).}
    \label{fig:tide_stations}
\end{figure}

Hourly sea-level observations are collected for six tide-gauge stations with distinct coastal environments: Venice and Trieste (Italy), Saint-Malo (France), Vard{\o} (Norway), Nikiski (Alaska, USA), and Nagasaki (Japan). Data for Venice are obtained from the Municipality of Venice \citep{comunevenezia2026}, those for Saint-Malo from the Hydrographic and Oceanographic Service of the French Navy \citep{shom_sealevel_2026}, and those for Trieste, Vard{\o}, Nikiski, and Nagasaki from the University of Hawaii Sea Level Center \citep{Caldwell2015}. The geographical location of the six stations is shown by the red dots in Figure~\ref{fig:tide_stations}.

\begin{figure}[H]
    \centering
    \includegraphics[width=0.90\textwidth]{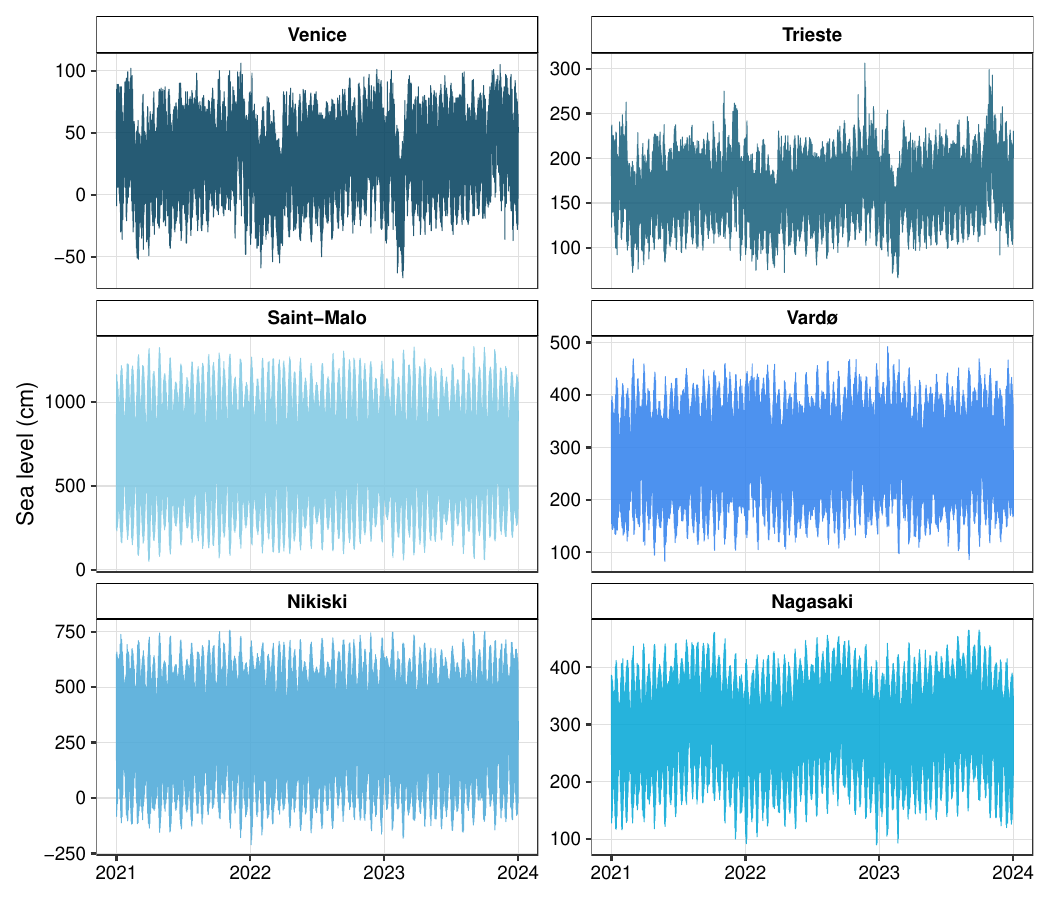}
    \caption{Hourly sea-level of the six tide-gauge stations over the period 2021--2023.}
    \label{fig:sea_level_series}
\end{figure}

\noindent
Venice is located within the Venice Lagoon, at the northern end of the Adriatic Sea. The lagoon is a shallow and morphologically complex environment, with an average depth of $\approx1.5~m$. It is connected to the Adriatic Sea through the three inlets: Lido, Malamocco, and Chioggia. The local tidal regime is predominantly semidiurnal, but water levels can be amplified by meteorological forcing. In particular, south-easterly winds (Sirocco) and low atmospheric pressure contribute to elevated sea levels, while north-easterly winds (Bora) may also affect water-level dynamics \citep{umgiesser2021}. These combined effects can generate high-water events locally known as \textit{``Aqua Alta''}, which have long affected the city's cultural heritage, environment, society, and economy. Flood exposure increases rapidly with water level: 90~cm above the local datum is sufficient to inundate the lowest areas, whereas levels exceeding 140~cm are considered exceptional (\textit{``Aqua Granda''}) and may flood nearly 60\% of the pedestrian area. The most recent major high-water event reached 189~cm on 12 November 2019, just below the record of 194~cm observed on 4 November 1966 \citep{umgiesser2021}. Trieste is located at the northeastern end of the Gulf of Trieste, in the northernmost part of the Adriatic Sea. Tidal oscillations are relatively large compared with most of the Mediterranean because the tidal signal is progressively amplified along the Adriatic basin toward the north. In the Gulf of Trieste, the astronomical tide is predominantly semidiurnal \citep{umgiesser2021}. Saint-Malo is situated on the northern coast of Brittany, within the macro-tidal Gulf of Saint-Malo and the broader Bay of Mont-Saint-Michel \citep{SHOM_GolfeNormandBreton}. Saint-Malo is included among the locations with the largest tidal range reported by \cite{NOAATidalRange}. Vard{\o} is located on a small island off the northeastern coast of Norway, facing the Barents Sea \citep{KartverketVardo}. Nikiski is located on the eastern shore of Cook Inlet, another strongly macro-tidal environment \citep{NOAATidalRange}. The morphology of the inlet contributes to a marked amplification of the tidal signal toward its upper reaches. Nagasaki is located within the deeply indented Nagasaki Bay on the western coast of Kyushu, Japan \citep{JMA_Nagasaki}. Figure~\ref{fig:sea_level_series} shows the hourly sea-level series for the six tide-gauge stations over the period 2021--2023. The figure highlights the strong heterogeneity in sea-level regimes. 

\noindent
Table~\ref{tab:descriptive_statistics} reports summary statistics for hourly sea-level observations and sea-level pressure at the six locations over the period January 2021--December 2023. The locations differ substantially in their average sea levels. Saint-Malo records the highest mean level (684.20~cm), followed by Nikiski (332.35~cm), Nagasaki (302.78~cm), and Vard{\o} (287.25~cm). Lower mean levels are observed at Trieste and Venice, with averages of 169.12~cm and 32.64~cm, respectively. The locations also display marked differences in variability. Saint-Malo is the most variable location ($\pm$ 282.88~cm), followed by Nikiski ($\pm$ 196.20~cm). Vard{\o} and Nagasaki show intermediate variability ($\pm$ 78.10~cm and $\pm$ 69~cm), while Trieste and Venice exhibit much smaller fluctuations ($\pm$ 31.15~cm and $\pm$ 28.20~cm). 

\begin{table}[H]
\centering
\caption{Descriptive statistics of hourly sea level (cm) and sea-level pressure (hPa) at the six locations over the period January 2021--December 2023.}
\label{tab:descriptive_statistics}
\begin{tabular}{lcccccccc}
\toprule
& \multicolumn{4}{c}{Sea level (cm)}
& \multicolumn{4}{c}{Sea-level pressure (hPa)} \\
\cmidrule(lr){2-5}\cmidrule(lr){6-9}
Location
& Mean & SD & Min & Max
& Mean & SD & Min & Max \\
\midrule
Venice
& 32.64 & 28.20 & -67.00 & 106.00
& 1016.32 & 7.95 & 983.20 & 1040.80 \\

Trieste
& 169.12 & 31.15 & 66.40 & 306.20
& 1015.94 & 7.72 & 984.60 & 1040.80 \\

Saint-Malo
& 684.20 & 282.88 & 50.30 & 1331.20
& 1017.30 & 9.72 & 967.20 & 1045.00 \\

Vard{\o}
& 287.25 & 78.10 & 82.70 & 492.10
& 1008.30 & 11.58 & 962.60 & 1042.50 \\

Nikiski
& 332.35 & 196.20 & -211.20 & 756.90
& 1008.10 & 12.21 & 960.10 & 1042.40 \\

Nagasaki
& 302.78 & 68.99 & 90.00 & 465.00
& 1015.68 & 7.28 & 967.80 & 1036.20 \\
\bottomrule
\end{tabular}
\end{table}

\subsection{Meteorological data}
The meteorological variables considered in the analysis are wind speed, wind direction, and sea-level atmospheric pressure. Meteorological observations at hourly frequency were obtained from the National Oceanic and Atmospheric Administration (NOAA) database using the \texttt{climate} package in \texttt{R} \citep{climate2020}. The nearest meteorological station was selected for each location. The wind is represented through its horizontal components rather than through wind speed and direction separately
\citep{Faranda2023,ecmwf2024wind}. Let $w_t$ denote wind speed, measured in $m/s$, and let $\theta_t$ denote wind direction, originally measured in degrees clockwise from North and subsequently converted to radians. Following the meteorological convention, wind direction indicates the direction from which the wind originates. The horizontal wind components (in $m/s$) are computed as $u_t = - w_t \sin(\theta_t)$ and $v_t = - w_t \cos(\theta_t)$. The component $u_t$ represents the east--west wind contribution, with positive values indicating eastward flow and negative values westward flow, whereas $v_t$ represents the north--south contribution, with positive values indicating northward flow and
negative values southward flow. This representation integrates both wind intensity and directional information. The dominant wind regimes differ across stations. Venice and Trieste are both characterized by a strong north-easterly component, accounting for 23.9\% and 34.6\% of hourly observations, respectively. In the northern Adriatic, this sector is associated with Bora wind. Venice also shows a relevant contribution from south-easterly, and southerly winds. The south-easterly sector is associated with Sirocco wind. Saint-Malo exhibits a more heterogeneous pattern, with southerly and south-westerly winds being the most frequent (20\%). Vard{\o} is mainly characterized by south-westerly and westerly winds (19.5\%). Nikiski shows a pronounced north-easterly regime (31.8\%), followed by southerly and easterly winds. Nagasaki presents a more distributed directional structure with northerly winds being the most frequent (22.1\%). Let $p_t$ denote sea-level atmospheric pressure, measured in hectopascal (hPa). During the period 2021--2023 (See Table~\ref{tab:descriptive_statistics}), Venice, Trieste, and Nagasaki exhibit similar pressure regimes, with mean values of 1016.32, 1015.94, and 1015.68~hPa, respectively. Saint-Malo records the highest mean pressure (1017.30~hPa), whereas Vard{\o} and Nikiski are characterized by lower mean values, 1008.30 and 1008.10~hPa, respectively, and the greatest variability, with standard deviations of 11.58 and 12.21~hPa. For Venice, operational forecasts of the wind components and sea-level pressure are obtained from NOAA's Global Forecast System (GFS), a global numerical weather prediction model that provides forecasts up to 16 days ahead \citep{noaa_gfs_aws}. GFS forecasts are issued four times daily, at three-hour intervals on a regular spatial grid. Assuming that each forecast becomes available six hours after initialization, the latest available forecast is selected at each forecast origin. Data are spatially and temporally interpolated to Venice with an hourly frequency. In 2023, the GFS values used to produce the 24-hour (48-hour) sea-level forecasts have correlations with the corresponding NOAA observations described above of 0.995 (0.988) for sea-level pressure and 0.63--0.67 (0.58--0.64) for the wind components.

\subsection{Site-specific operational data}
For Venice, a binary covariate is introduced to account for the effect of MoSE on observed sea levels. Specifically, $\text{MoSE}_{t}$ is defined
as
\begin{equation}
    \text{MoSE}_{t} =
    \begin{cases}
        1, & \text{if the three MoSE flood-barriers are activated at time } t,\\
        0, & \text{otherwise.}
    \end{cases}
\end{equation}

\section{Methodology}\label{sec:method}
The observed sea level at time $t$, denoted by $Y_t$, is decomposed into an astronomical component $A_t$ and a non-astronomical component $N_t$:
\begin{equation}
    Y_t = A_t + N_t.
    \label{eq:sea_level_decomposition}
\end{equation}
While $A_t$ is estimated through HA, the non-astronomical component is modelled using a set of statistical and data-driven approaches. In general, the non-astronomical component is specified as
\begin{equation}
    N_t = f(X_t) + \varepsilon_t,
    \label{eq:nonastronomical_component}
\end{equation}
where $X_t$ denotes the information set available at time $t$, $f(\cdot)$ represents the systematic component captured by the model, and $\varepsilon_t$ is the part of the non-astronomical component that remains unexplained.\\ 
The information set includes the meteorological covariates, relevant lagged values of $N_t$, and site-specific operational variables accounting for external interventions or local operating conditions. Combining Eqs.~\eqref{eq:sea_level_decomposition} and
\eqref{eq:nonastronomical_component}, the observed sea level is therefore specified as
\begin{equation}
    Y_t = A_t + f(X_t) + \varepsilon_t.
    \label{eq:sea_level_model}
\end{equation}

\subsection{Astronomical component modelling}
The astronomical component $A_t$ is estimated through HA \citep{Godin1972,PAWLOWICZ2002,franco2009mares,stephenson2016,abubakar2019}. The method represents the astronomical tide as the superposition of a finite number of sinusoidal constituents as follows:
\begin{equation}
    A_t =
    \mu +
    \sum_{k=1}^{K}
    H_k F_k(t)
    \cos\left[
    \frac{\pi}{180}
    \left(
    \omega_k t - g_k + u_k(t) + V_k
    \right)
    \right],
    \label{eq:harmonic_component}
\end{equation}
where $\mu$ is the mean sea level and $K$ is the number of tidal constituents included in the model. For each $k$th constituent, $\omega_k$ is the known astronomical angular speed, whereas $H_k$ and $g_k$ are unknown parameters representing the amplitude and phase lag, respectively. The latter is defined relative to the equilibrium tide at the Greenwich meridian \citep{pugh1987tides}. The terms $F_k(t)$ and $u_k(t)$ denote the known nodal corrections to amplitude and phase, respectively. They are designed to capture long-period astronomical cycles. The term $V_k$ is the known astronomical argument defined with respect to the selected reference epoch. Thus, $\omega_k$, $F_k(t)$, $u_k(t)$, and $V_k$ are known astronomical quantities, whereas $\mu$, $H_k$, and $g_k$ must be estimated from the observed sea-level series. The unknown harmonic parameters are estimated using the least-squares method, which identifies the amplitudes and phases that minimize the squared differences between observed and reconstructed water levels. The mean sea level $\mu$ is first computed as the sample mean and removed from the series. Each tidal constituent is then expressed as a linear combination of sine and cosine terms, allowing the corresponding coefficients to be obtained through zero-intercept linear regression. Sine and cosine coefficients are subsequently converted into the amplitude $H_k$ and phase lag $g_k$ of each constituent. A model with $K$ tidal constituents requires the estimation of $2K+1$ parameters. HA is  performed using the \texttt{R} package \texttt{TideHarmonics} \citep{stephenson2016}. A key modelling choice is the selection of the $K$ constituents. \cite{Doodson1921} identified 388 tidal constituents \citep{doodson1954,CasottoBiscani2004}. However, in many cases a substantially smaller subset is sufficient for practical tide predictions \citep{Marone2013,yin2015hybrid,stephenson2016,abubakar2019,jiao2026}. In the present analysis, the astronomical component is represented by eight major tidal constituents \citep{ISPRA2012,Okwuashi2017,ElDiasty2018,abubakar2019,Liu2019,dinunno2021,ComuneVeneziaConst,Monahan2023}. The semi-diurnal constituents are the principal lunar constituent $\mathrm{M2}$ (period of 12.42 hours), the principal solar constituent $\mathrm{S2}$ (12.00 hours), the larger lunar elliptic constituent $\mathrm{N2}$ (12.66 hours), and the lunisolar constituent $\mathrm{K2}$ (11.97 hours). The diurnal constituents comprise the lunisolar constituent $\mathrm{K1}$ (23.93 hours), the principal lunar constituent $\mathrm{O1}$ (25.82 hours), the principal solar constituent $\mathrm{P1}$ (24.07 hours), and the solar constituent $\mathrm{S1}$ (24.00 hours). Setting $K=8$, the resulting harmonic model contains 17 parameters: the mean sea level and two parameters for each tidal constituent.

\subsection{Non-astronomical component modelling}
\begin{figure}[H]
    \centering
    \includegraphics[width=0.90\textwidth]{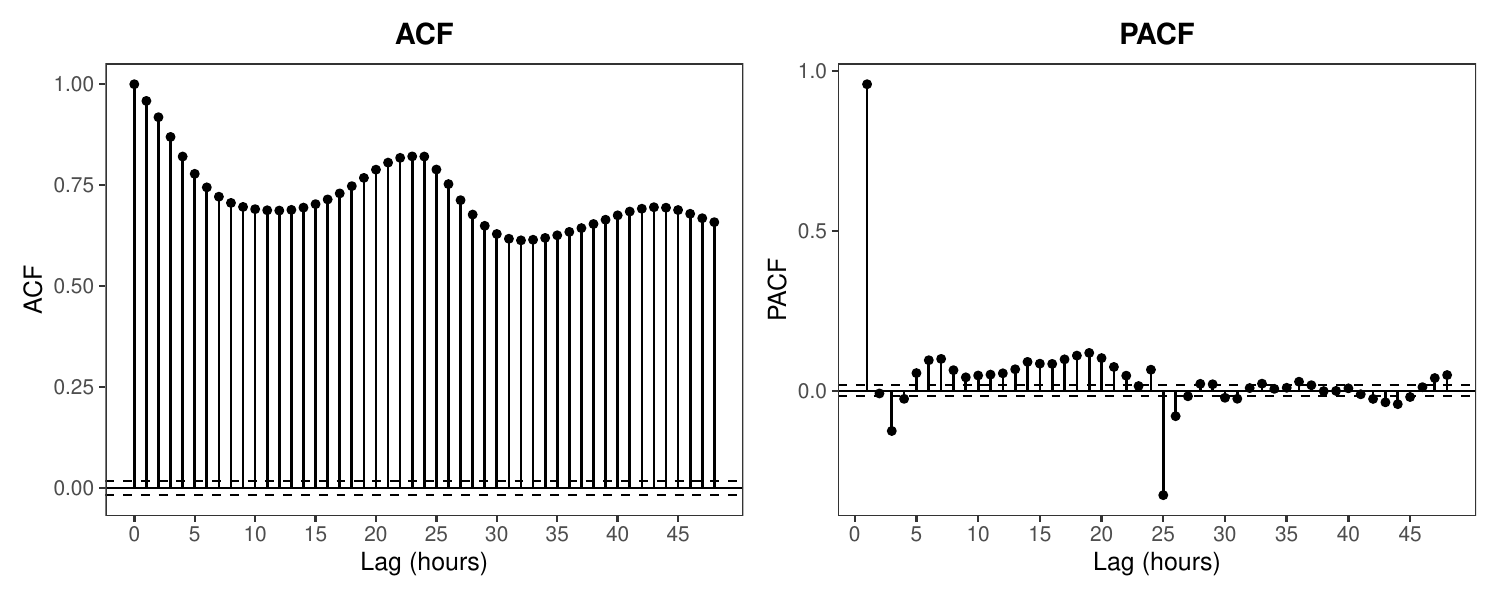}
    \caption{Autocorrelation function (ACF, left) and partial autocorrelation function (PACF, right) of the estimated non-astronomical component for Venice.}
    \label{fig:venice_acf_pacf}
\end{figure}

As a first specification, the function $f(X_t)$ in Eq.~\eqref{eq:nonastronomical_component} is linearly modelled using an autoregressive model with exogenous covariates (ARX):
\begin{equation}
\begin{aligned}
    N_t
    &=
    \beta_0
    + \beta_1 u_{t-1}
    + \beta_2 v_{t-1}
    + \beta_3 uv_{t-1}
    + \beta_4 p_{t-1}
    + \beta_5 \text{MoSE}_{t-1}
    + \sum_{\ell \in \mathcal{L}} \phi_{\ell} N_{t-\ell}
    + \epsilon_t,
\end{aligned}
\label{eq:linear_residual_model}
\end{equation}
where $N_t=Y_t-A_t$ is the non-astronomical residual component at time $t$, $u_{t-1}$ and $v_{t-1}$ are the one-hour lagged wind components, and $p_{t-1}$ is the one-hour lagged sea-level pressure. The interaction term $uv_{t-1}$ allows the effect of wind forcing to depend jointly on the two wind components. The term $\epsilon_t$ represents the unexplained variation in the residual process. The vector of predictors is defined as
$X_t = \bigl(
u_{t-1},\,\allowbreak
v_{t-1},\,\allowbreak
uv_{t-1},\,\allowbreak
p_{t-1},\,\allowbreak
\text{MoSE}_{t-1},\,\allowbreak
\{N_{t-\ell}\}_{\ell\in\mathcal{L}}
\bigr)$. Unless otherwise stated, the same specification of $X_t$ is used throughout the analysis. The variable $\text{MoSE}_{t-1}$ is included only for Venice. The autoregressive structure is defined over the lag set
$\mathcal{L} = \{1,2,3,4,20,21,22,23,24,25\}$. Equivalently, $\phi_\ell = 0$ for $\ell \notin \mathcal{L}$. This specification is designed to capture two distinct regions of serial dependence in the non-astronomical component: short-term persistence over the first few hours and residual dependence around the daily time scale. The choice of these lag blocks is guided by the empirical autocorrelation function (ACF) and partial autocorrelation function (PACF) of $\widehat{N}_t$, computed over the validation set. Figure~\ref{fig:venice_acf_pacf} reports the corresponding ACF and PACF for Venice. 
\noindent
The non-astronomical component exhibits strong persistence at short lags, together with a secondary dependence pattern emerging at approximately 20--48 hours. Rather than including all intermediate lags, only the lag blocks corresponding to these dominant dependence structures are retained, with the coefficients of all remaining lags constrained to zero. This yields a more parsimonious autoregressive specification while preserving the main serial dependence patterns.

\begin{figure}[H]
    \centering
    \includegraphics[width=0.90\textwidth]{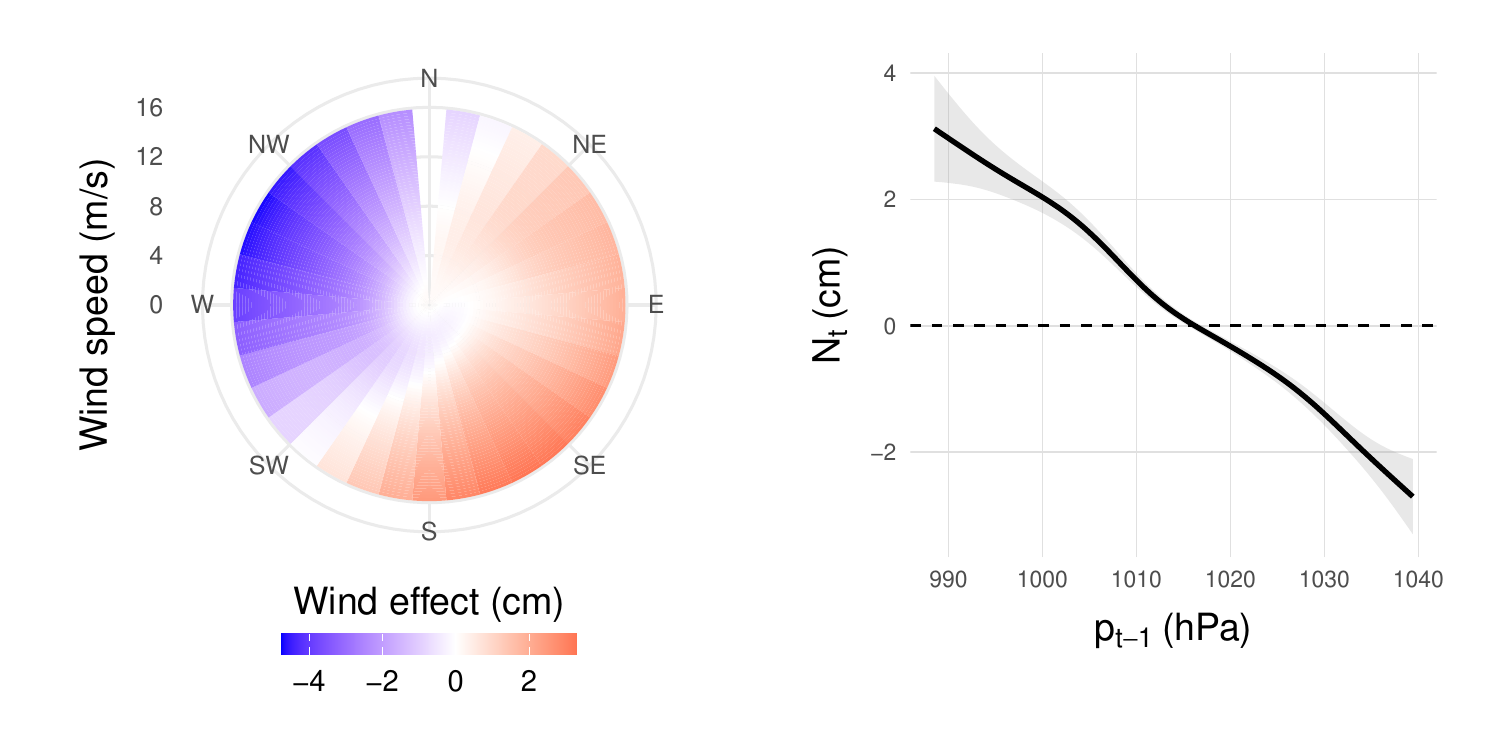}
    \caption{Wind (left) and sea level pressure (right) effects estimated by the GAM model on the non-astronomical component for Venice in 2021-2023.}
    \label{fig:gam_effects}
\end{figure}

As a second specification, the function $f(X_t)$ in
Eq.~\eqref{eq:nonastronomical_component} is modelled using a generalized additive model (GAM). The vector of predictors is defined as
$X_t = \bigl(
u_{t-1},\,\allowbreak
v_{t-1},\,\allowbreak
p_{t-1},\,\allowbreak
\text{MoSE}_{t-1},\,\allowbreak
\{N_{t-\ell}\}_{\ell\in\mathcal{L}}
\bigr)$. The specification preserves the autoregressive structure of the ARX model while allowing the effects of the meteorological covariates to enter nonlinearly:
\begin{equation}
\begin{aligned}
    N_t
    &=
    \beta_0
    + \beta_1 \text{MoSE}_{t-1}
    + s_1\!\left(u_{t-1},v_{t-1}\right)
    + s_2\!\left(p_{t-1}\right)
    + \sum_{\ell \in \mathcal{L}}
        \phi_{\ell} N_{t-\ell}
    + \epsilon_t ,
\end{aligned}
\label{eq:gam_residual_model}
\end{equation}
where $s_1(\cdot,\cdot)$ is a bivariate smooth function of the one-hour lagged wind components and $s_2(\cdot)$ is a univariate smooth function of the one-hour lagged mean sea-level pressure. The term $\text{MoSE}_{t-1}$ and autoregressive terms enter linearly. The bivariate smooth $s_1(u_{t-1},v_{t-1})$ allows the effect of wind forcing on the non-astronomical component to vary flexibly with both wind magnitude and direction, without imposing a predefined functional form. Similarly, $s_2(p_{t-1})$ accommodates a nonlinear response to atmospheric pressure. The smooth functions are represented using penalized regression spline bases, with basis dimensions $k=20$ for $s_1(u_{t-1},v_{t-1})$ and $k=5$ for $s_2(p_{t-1})$. These values determine the maximum flexibility available to the corresponding smooth terms, whereas their effective complexity is controlled by the associated smoothness penalties. The GAM is estimated in \textsf{R} using the \texttt{mgcv} package \citep{Wood2017}. Given the Gaussian response specification, the parametric coefficients and spline coefficients are estimated jointly by penalized least squares, while the smoothing parameters controlling the effective complexity of the smooth terms are selected by restricted maximum likelihood (REML). An interesting feature of the GAM model is the interpretability of the estimated meteorological effects. Figure~\ref{fig:gam_effects} illustrates the partial effects of lagged wind and sea-level pressure on the non-astronomical component for Venice during 2021-2023. 

\noindent
The left panel represents the bivariate smooth $s(u_{t-1},v_{t-1})$ in polar coordinates, where the radius denotes wind speed and the angle indicates the wind direction of origin. A clear directional pattern emerges. North-easterly (NE), easterly (E) and south-easterly (SE) winds, are associated with positive contributions to the non-astronomical component. At the wind-speed $>10 m/s$, south-easterly winds (Sirocco) produce the strongest positive effect, increasing the non-astronomical component by about 2.5 cm on average. North-easterly winds (Bora) also show a positive effect, of about 1.4 cm on average. By contrast, westerly (W) and north-westerly (NW) winds produce negative effects. These winds flush water out of the lagoon. North-easterly (NE) and south-easterly (SE) winds are among the most frequent wind directions in Venice. The right panel shows a strongly decreasing effect of lagged sea-level pressure. Low-pressure conditions are associated with positive contributions to the non-astronomical component, whereas high-pressure conditions reduce it. The effect crosses zero at approximately 1015--1020~hPa. The function $s_2(p_{t-1})$ captures the inverse-barometer effect \citep{dinunno2021,umgiesser2021,delauro2023}.

\begin{figure}[H]
\centering

\begin{tikzpicture}[
    scale=1,
    transform shape,
    node distance=0.60cm and 1.20cm,
    input/.style={
        rectangle,
        rounded corners=2pt,
        draw,
        minimum width=2.20cm,
        minimum height=0.55cm,
        inner sep=2pt,
        fill=blue!6,
        font=\scriptsize
    },
    hidden/.style={
        circle,
        draw,
        minimum size=0.68cm,
        inner sep=0pt,
        fill=cyan!12
    },
    output/.style={
        circle,
        draw,
        minimum size=0.80cm,
        inner sep=0pt,
        fill=orange!18
    },
    bias/.style={
        circle,
        draw,
        minimum size=0.42cm,
        inner sep=0pt,
        fill=gray!15,
        font=\scriptsize
    },
    arr/.style={
        -{Latex[length=1.8mm]},
        line width=0.3pt
    },
    lab/.style={
        font=\scriptsize
    }
]

\node[input] (z1) {$u_{t-1}$};
\node[input, below=of z1] (z2) {$v_{t-1}$};
\node[input, below=of z2] (z3) {$uv_{t-1}$};
\node[input, below=of z3] (z4) {$p_{t-1}$};
\node[input, below=of z4] (z5) {$\text{MoSE}_{t-1}$};

\node[input, below=0.80cm of z5] (z6)
{$\{N_{t-\ell}\}_{\ell\in\mathcal{L}}$};

\node[hidden, right=1.75cm of z2] (h1) {$h_1$};
\node[hidden, below=0.70cm of h1] (h2) {$h_2$};
\node[below=-0.15cm of h1] (hdots) {$\vdots$};
\node[below=0.33cm of h2] (hdots) {$\vdots$};
\node[hidden, below=0.33cm of hdots] (hq) {$h_q$};

\node[output, right=1.75cm of h2] (out)
{$\widehat{N}_t$};

\foreach \i in {z1,z2,z3,z4,z5,z6} {
    \draw[arr] (\i.east) -- (h1.west);
    \draw[arr] (\i.east) -- (h2.west);
    \draw[arr] (\i.east) -- (hq.west);
}

\draw[arr] (h1.east) -- (out.west);
\draw[arr] (h2.east) -- (out.west);
\draw[arr] (hq.east) -- (out.west);


\node[lab, above=0.20cm of z1] {\textbf{Input layer}};
\node[lab, above=0.20cm of h1] {\textbf{Hidden layer}};
\node[lab, above=0.20cm of out] {\textbf{Output layer}};

\node[lab] at ($(z3)!0.52!(h2) + (0,2.3)$)
{$\boldsymbol{\gamma}_j$};

\node[lab] at ($(h2)!0.50!(out) + (0,0.34)$)
{$\nu_j$};

\node[lab, right=0.10cm of hq]
{$\sigma(\cdot)$};

\end{tikzpicture}

\caption{Representation of the NNARX model for the non-astronomical component.}
\label{fig:nnarx_architecture}
\end{figure}
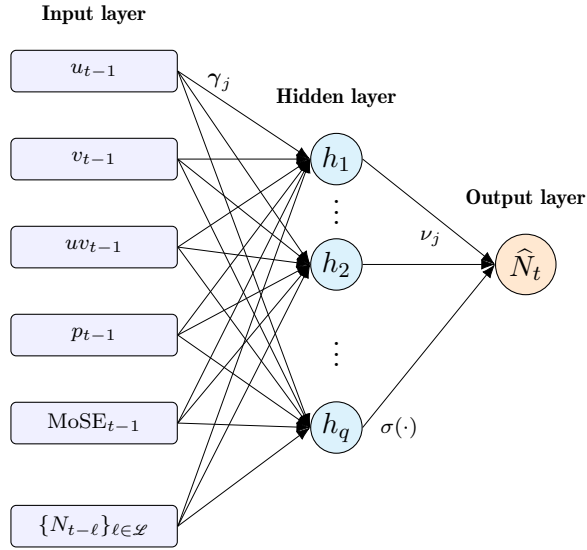

As a further nonlinear specification, the function $f(X_t)$ is modelled using an autoregressive (feed-forward) neural network with exogenous inputs. Unlike the ARX specification, the NNARX model does not impose a linear relationship between the predictors and the non-astronomical component. For $q$ hidden neurons, the model can be expressed as
\begin{equation}
    N_t
    =
    \nu_0
    +
    \sum_{j=1}^{q}
    \nu_j
    \sigma\!\left(
        \gamma_{0j}
        +
        \boldsymbol{\gamma}_j^{\top}X_t
    \right)
    +
    \epsilon_t,
    \label{eq:mlp_residual}
\end{equation}
where $\sigma(\cdot)$ denotes the hidden-layer activation function, $\boldsymbol{\gamma}_j$ is the vector of weights connecting the input layer to the $j$th hidden neuron, $\gamma_{0j}$ is its bias term,
$\nu_j$ is the weight connecting the $j$th hidden neuron to the output unit, and $\nu_0$ is the output-layer bias. Figure \ref{fig:nnarx_architecture} represents the NNARX architecture for the non-astronomical component. 

The number of hidden neurons $q$ is treated as a tuning parameter and selected on the validation set according to out-of-sample predictive performance, as measured by the mean absolute error (MAE) and root mean squared error (RMSE). Three network sizes are considered, with $q \in \{3,8,14\}$, and the number of hidden neurons is selected separately for each station using the validation set. When competing architectures yield comparable validation accuracy, preference is given to the more parsimonious specification. The selected network size varies across locations: three hidden neurons are chosen for Venice, Trieste, Nikiski, and Nagasaki, whereas Saint-Malo and Vard{\o} require 8 and 14 hidden neurons, respectively.

\begin{figure}[H]
    \centering
    \includegraphics[width=0.90\textwidth]{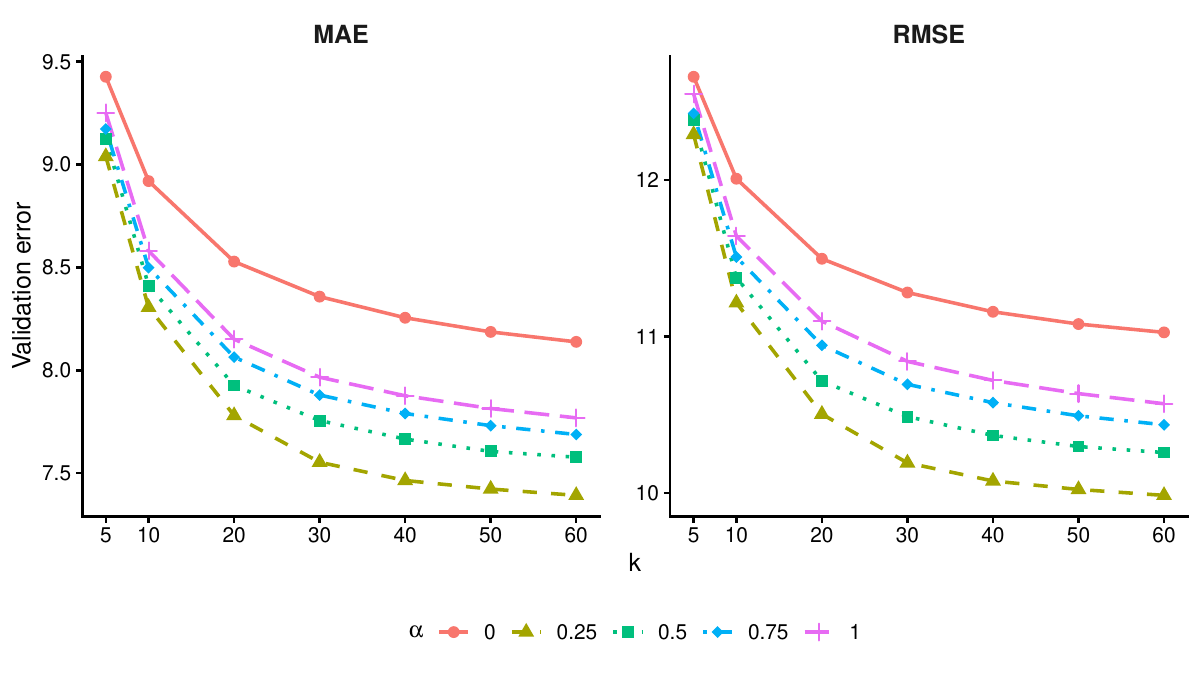}
    \caption{Validation-based selection of the KNN hyperparameters $\alpha$ and $k$ for Venice. MAE (left) and RMSE (right).}
    \label{fig:knn_validation}
\end{figure}

As an additional non-parametric specification, the non-astronomical component is modelled using a non-standard $k$-nearest neighbours (KNN) approach. The method exploits similarities between daily profiles of the non-astronomical component and associated meteorological conditions, without imposing a predefined functional form. KNN methods have been applied to time-series forecasting by identifying historical patterns that are similar to the most recently observed trajectory and using their subsequent evolution to construct predictions \citep{martinez2019}. Recent formulations extend this idea to subsequence matching and weighted nearest-neighbour forecasting \citep{tajmouati2024}. Similarity measures can also incorporate exogenous information, allowing neighbour selection to account jointly for the recent behaviour of the target series and external predictors \citep{trupiano2021,Faranda2023}. Let $\mathbf{N}_i=(N_{i,1},\ldots,N_{i,24})^{\top}$ denote the 24-hour
profile of the non-astronomical component on day $i$, where $N_{i,h}$ is the residual component at hour $h$. To predict the profile for day $i$, the KNN procedure searches for historical daily profiles similar to the most recently observed profile $\mathbf{N}_{i-1}$. Similarity in the residual component between day $i-1$ and a generic historical day $j$ is measured by the Euclidean distance
\begin{equation}
    D_L(i-1,j)
    =
    \sqrt{
        \sum_{h=1}^{24}
        \left(
            N_{i-1,h}-N_{j,h}
        \right)^2
    }.
    \label{eq:knn_residual_distance}
\end{equation}
Neighbour selection also accounts for meteorological similarity through the wind-interaction term $u_{i,h}v_{i,h}$ and mean sea-level pressure
$p_{i,h}$. Both variables are standardized using the mean and standard deviation computed from the available historical sample. Denoting standardized values by a tilde, meteorological similarity between day $i-1$ and a generic historical day $j$ is measured as
\begin{equation}
D_M(i-1,j)
=
\sqrt{
\frac{1}{2}
\left\{
\sum_{h=1}^{24}
\left(
\widetilde{u_{i-1,h}v_{i,h}}
-
\widetilde{u_{j,h}v_{j,h}}
\right)^2
+
\sum_{h=1}^{24}
\left(
\widetilde{p}_{i-1,h}
-
\widetilde{p}_{j,h}
\right)^2
\right\}
}.
\label{eq:knn_meteo_distance}
\end{equation}
The factor $1/2$ assigns equal importance to the two standardized meteorological variables. This distance construction follows the approach of \cite{Faranda2023}, which combines multiple meteorological variables into a common Euclidean distance with equal weights. The two sources of similarity are then combined through
\begin{equation}
    D(i,j)
    =
    \sqrt{
        \frac{1}{24}
        \bigg[
            \alpha D_L(i-1,j)^2
            +
            (1-\alpha)D_M(i-1,j)^2
        \bigg]
    },
    \qquad
    \alpha\in[0,1],
    \label{eq:knn_total_distance}
\end{equation}
where $\alpha$ controls the relative importance assigned to the recent non-astronomical trajectory and meteorological conditions. The factor $1/24$ averages the squared discrepancies over the 24 hourly observations. Let $j_1,\ldots,j_k$ denote the $k$ historical days with the smallest combined distances $D(i,j)$, which jointly reflect residual and meteorological similarity. The forecast for day $i$ is obtained as
\begin{equation}
    \widehat{\mathbf{N}}_i
    =
    \sum_{r=1}^{k}
    w_r\,\mathbf{N}_{j_r+1},
\end{equation}
where the rank-order centroid weights
\begin{equation}
    w_r
    =
    \frac{1}{k}
    \sum_{m=r}^{k}\frac{1}{m},
    \qquad r=1,\ldots,k,
\end{equation}
assign greater importance to closer neighbours. The hyperparameters $k$ and $\alpha$ are selected through a grid search on a separate validation set. The candidate values are
\[
k \in \{5,10,20,30,40,50,60\},
\qquad
\alpha \in \{0,0.25,0.50,0.75,1\}.
\]
The limiting cases $\alpha=0$ and $\alpha=1$ correspond to two special specifications: for $\alpha=0$, neighbour selection is based exclusively
on meteorological similarity, whereas for $\alpha=1$, it is based exclusively on the recent non-astronomical trajectory. Intermediate values of $\alpha$ combine the two sources of information. For example, $\alpha=0.50$ assigns equal weights to $D_L$ and $D_M$. Model performance over the grid is evaluated minimizing MAE and RMSE. The elbow method is also considered to avoid unnecessarily complex specifications when additional neighbours provide only marginal improvements. For Venice, Figure~\ref{fig:knn_validation} summarizes the joint tuning of $k$ and $\alpha$. 
\noindent
Validation errors decrease as $k$ increases and then gradually level off, supporting the use of the elbow criterion. The limiting cases $\alpha=0$ and $\alpha=1$, corresponding respectively to purely meteorological and purely non-astronomical neighbour selection, are generally outperformed by intermediate values of $\alpha$. The validation-based selection of the KNN hyperparameters reveals clear differences across stations. Venice, Trieste, and Nagasaki select $k=20$ and $\alpha=0.25$, while Saint-Malo selects $k=10$ and $\alpha=0.25$. By contrast, Vard{\o} and Nikiski select $\alpha=1$, with $k=20$ and $k=10$, respectively, thereby relying exclusively on $N_t$. The optimal value of $\alpha$ reflects the predictive contribution of the meteorological variables and may also depend on the geographical setting. This is particularly evident for the geographically close stations of Venice and Trieste, which select the same hyperparameter combination.

\begin{figure}[H]
    \centering
    \includegraphics[width=0.90\textwidth]{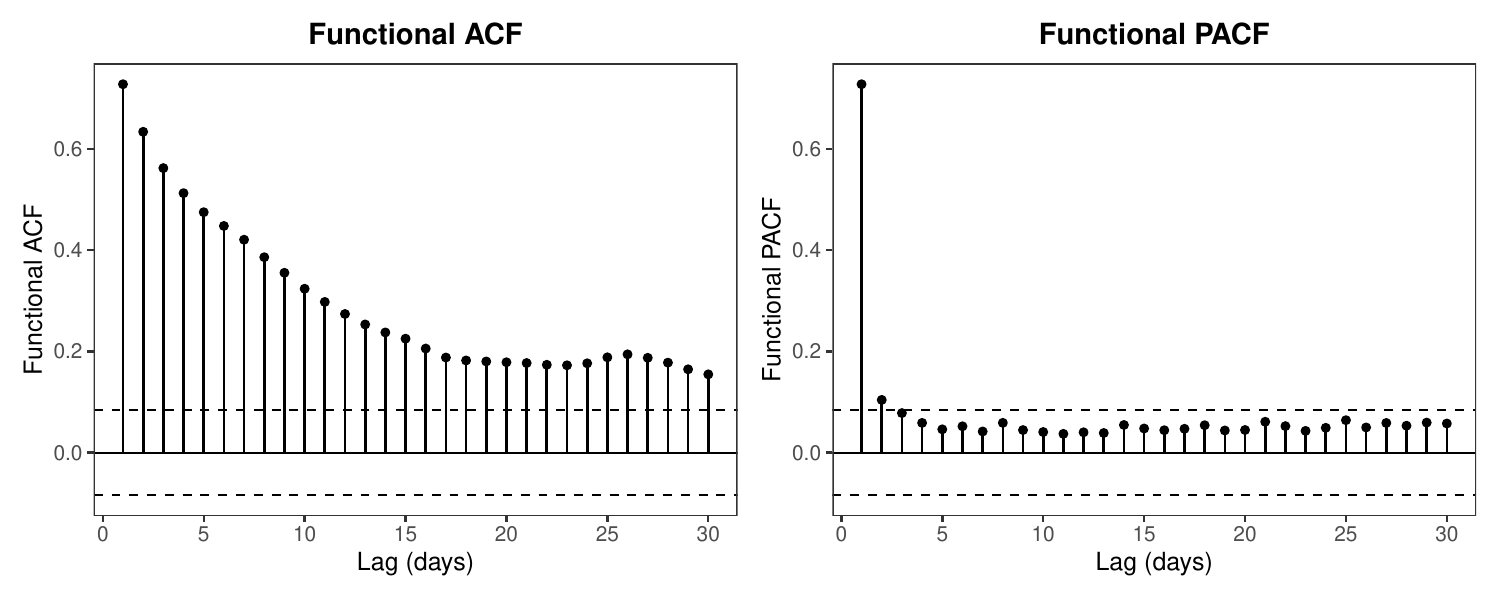}
    \caption{Functional autocorrelation function (fACF, left) and functional partial autocorrelation function (fPACF, right) of the estimated non-astronomical component for Venice over lags of up to 30 days.}
    \label{fig:venice_Rt_functional_acf_pacf}
\end{figure}

As a functional extension of the previous specifications, the non-astronomical component can be modelled by a functional autoregressive model with exogenous covariates (FARX). In this framework, the hourly observations are represented as daily functional profiles. Let $\mathcal{H}$ be a real separable Hilbert space and consider the non-astronomical component $N_i(\tau)\in L^2[1,24]$, where $i$ indexes the daily profiles and $\tau\in[1,24]$ denotes the hour within the day \citep{ramsay2006}. The vector of predictors is defined as
$X_t = \bigl(
u_{i-1}(\tau),\,\allowbreak
v_{i-1}(\tau),\,\allowbreak
uv_{i-1}(\tau),\,\allowbreak
p_{i-1}(\tau),\,\allowbreak
\text{MoSE}_{i-1}(\tau),\,\allowbreak
N_{i-1}
\bigr)$. A FARX(1) specification is considered, in which the current non-astronomical profile depends on the profile observed on the previous day and on the corresponding lagged exogenous functional covariates \citep{Damon2002}:
\begin{equation}
\begin{aligned}
N_i(\tau)
=&
\rho\!\left(N_{i-1}(\tau)\right)
+\psi_1\!\left(MoSE_{i-1}(\tau)\right)
+\psi_2\!\left(u_{i-1}(\tau)\right)
+\psi_3\!\left(v_{i-1}(\tau)\right) \\
&+
\psi_4\!\left(uv_{i-1}(\tau)\right)
+\psi_5\!\left(p_{i-1}(\tau)\right)
+\epsilon_i(\tau),
\end{aligned}
\label{eq:farx1_residual}
\end{equation}
where $\rho(\cdot)$, $\psi_1(\cdot)$, $\psi_2(\cdot)$,
$\psi_3(\cdot)$, $\psi_4(\cdot)$, and $\psi_5(\cdot)$ are bounded linear operators on $\mathcal{H}$. The operator $\rho(\cdot)$ captures the dependence of the current non-astronomical profile on its previous-day profile. The remaining operators describe the effects of the lagged exogenous functional covariates. $MoSE_{i-1}(\tau)$,$u_{i-1}(\tau)$,$v_{i-1}(\tau)$,$uv_{i-1}(\tau)$, and $p_{i-1}(\tau)$ are continuous exogenous variables defined in $\mathcal{H}$. The term $\epsilon_i(\tau)$ denotes a functional white-noise process in $\mathcal{H}$. The functional representation preserves the within-day evolution of the non-astronomical component and of the exogenous variables, while allowing dependence between consecutive daily profiles to be modelled through linear operators. The number of discretization points is fixed at 24, corresponding to the 24 hourly observations defining each daily profile. The process order is fixed at one based on the functional autocorrelation function (fACF) and functional partial autocorrelation function (fPACF) computed on $N_i(\tau)$ over the validation set \citep{fdaACF,Mestre2021}. Figure~\ref{fig:venice_Rt_functional_acf_pacf} reports the fACF and fPACF of $\widehat{N}_t$ for Venice over lags of up to 30 days. While the fACF exhibits a gradual decay across successive daily lags, the fPACF is dominated by a pronounced spike at lag one, with substantially smaller partial correlations at higher lags. This pattern is consistent with a first-order functional autoregressive dependence structure and motivates the use of a FARX(1) specification. Similar patterns are observed for the other tide-gauge stations. The FARX(1) model is estimated in \textsf{R} using the \texttt{far} package \citep{Damon2024}.

\section{Forecasting design}
\label{sec:forecasting_design}
The dataset is divided into three subsets: training, validation, and test. The training set spans from 1 January 2021 to 30 June 2022, the validation set from 1 July 2022 to 31 December 2022, and the test set from 1 January 2023 to 31 December 2023. The validation set is used to tune the hyperparameters of the NNARX and KNN models, while the test set is used to assess the out-of-sample forecasting performance of all competing models. Forecasts are generated using a rolling-origin scheme with an expanding window. At each forecast origin, predictions are produced for horizons $h=1,\ldots,120$, corresponding to one hour to five days ahead. The forecast origin is then advanced by one hour, and the procedure is repeated throughout the test set, producing 8,760 forecasts for each horizon $h$. At each forecast origin, the astronomical component is first estimated using HA. The non-astronomical component is then computed over the available historical sample as the difference between the observed sea level and its harmonic estimate, $\widehat{N}_t = Y_t-\widehat{A}_t.$
A data-driven model is fitted to this residual component and used to generate the forecast $\widehat{N}_{t+h}$. The final sea-level forecast is obtained by combining the forecasts of the astronomical and non-astronomical components $\widehat{Y}_{t+h}=\widehat{A}_{t+h}+\widehat{N}_{t+h}$. The forecasting exercise considers five hybrid specifications, denoted by \mbox{HA+ARX}, \mbox{HA+GAM}, \mbox{HA+NNARX}, \mbox{HA+KNN}, and \mbox{HA+FARX}, each combining harmonic analysis with one of the previously introduced data-driven models. The models are re-estimated at each forecast origin to update both components.\\ 
In the main experiment, forecasts are generated using true, and thus not yet observed meteorological covariates. This design gives all models the same covariate information and excludes errors in the meteorological forecasts, which would otherwise propagate into the sea-level forecasts \citep{hyndman2021fpp,umgiesser2021}. To assess how the discrepancy between forecast and observed meteorological conditions affects sea-level forecast accuracy, the analysis is further extended to a real-time setting. Meteorological observations are used up to the last observed hour, values beyond that point are taken from the most recent GFS forecast available at the forecast origin. Since this forecast was issued 6 to 17 hours earlier, its inputs have lead times of 29--40 hours for the 24-hour sea-level forecasts and 53--64 hours for the 48-hour forecasts. Models specification and estimation procedures remain unchanged. The 24-hour forecasts from HA+FARX and HA+KNN are unchanged, as these models do not require meteorological values after the forecast origin at this horizon. This exercise, carried out for Venice as a case study and the cost--loss analysis in Section~7 is based on GFS inputs.

Forecast accuracy is evaluated using both MAE and RMSE. For each horizon $h$, the MAE is defined as follows:
\begin{equation}
\mathrm{MAE}_h
=
\frac{1}{T_h}
\sum_{i=1}^{T_h}
\left|
Y_{i,h}-\widehat{Y}_{i,h}
\right|,
\qquad h=1,\ldots,120,
\end{equation}
where $Y_{i,h}$ and $\widehat{Y}_{i,h}$ denote the observed and predicted sea levels, respectively, for forecast origin $i$ at horizon $h$, and $T_h$ is the number of available forecasts at that horizon. For each horizon $h$, the RMSE is defined as follows:
\begin{equation}
\mathrm{RMSE}_h
=
\sqrt{
\frac{1}{T_h}
\sum_{i=1}^{T_h}
\left(
Y_{i,h}-\widehat{Y}_{i,h}
\right)^2
},
\qquad h=1,\ldots,120.
\end{equation}

The Diebold--Mariano (DM) test is used to assess whether differences in forecast accuracy between competing models are statistically significant \citep{Diebold1995}. The test was originally developed to compare predictive accuracy rather than to perform structural model selection, and caution has therefore been raised against its mechanical interpretation in pseudo-out-of-sample model comparisons \citep{Diebold2015}. In the present study, the DM test is used exclusively as a formal comparison of the predictive performance of the competing hybrid models. Let $e_{a,t}$ and $e_{b,t}$ denote the forecast errors of models $a$ and $b$, respectively. Under absolute-error loss, the loss differential is defined as
$d_{t,ab}
=
\left|e_{a,t}\right|
-
\left|e_{b,t}\right|$.
Hence, $d_{t,ab}>0$ indicates that model $b$ has a smaller absolute forecast error than model $a$. The one-sided DM test is therefore based on
\begin{equation}
\begin{cases}
H_0: \mathbb{E}(d_{t,ab}) = 0\\
H_1: \mathbb{E}(d_{t,ab}) > 0
\end{cases}
\end{equation}
where the alternative hypothesis corresponds to model $b$ having superior predictive accuracy. The DM test statistic is computed as follows:
\begin{equation}
\label{eq:dmtest}
DM_{ab}
=
\frac{\bar{d}_{ab}}{\sqrt{(1/T) \sum_{k=-m}^{m} \gamma_k}},
\end{equation}
where: $\bar{d}_{ab}$ is the sample mean of the loss differential $d_{t,ab}$. The term $\gamma_k$ denotes the $k$-th order autocovariance of the loss differential series $d_{t,ab}$; $T$ is the number of forecast periods; $m$ is the maximum lag order used to estimate the long-run variance of $d_{t,ab}$. Under the null hypothesis, the DM statistic is asymptotically standard normal. Since multiple pairwise DM tests are performed across the competing models, a multiple-testing correction is required. Forecast errors are obtained from the same underlying sea-level series and are therefore potentially dependent. The Benjamini--Yekutieli (BY) procedure is applied to control the false discovery rate under general dependence among the test statistics \citep{BenjaminiYekutieli2001}. The DM test is complemented by the model confidence set (MCS). The MCS identifies a subset of models whose predictive performance cannot be statistically distinguished at a given confidence level \citep{Hansen2011}. Starting from an initial set of competing models $M_0$, the procedure sequentially tests the null hypothesis of equal predictive ability,
\begin{equation}
H_0:
\mathbb{E}(d_{t,ab})=0
\qquad
\forall\, a,b \in M,
\end{equation}
for the current set $M$. If the null is rejected, the model displaying the weakest relative predictive performance is removed and the test is repeated. The procedure continues until the null of equal predictive ability can no longer be rejected. The resulting set $\widehat{M}_{1-\alpha}^{*}$ is referred to as the superior set of models at confidence level $1-\alpha$. Because the distribution of the MCS test statistic is non-standard and forecast losses may be serially dependent, inference is obtained by bootstrap \citep{bernardi2014}.

Point forecasts provide a single predicted value but do not quantify uncertainty about future outcomes. This limitation is particularly relevant when extreme outcomes are costly or decision losses are asymmetric, as in the case of high sea levels. Probabilistic forecasts make this uncertainty explicit and support decision-making across a range of plausible scenarios \citep{taylor2026}. Since the main operational concern is the occurrence of high sea levels, the probabilistic evaluation focuses on the right tail of the predictive distribution.\\
\noindent
Quantile forecasts are obtained using the historical simulation approach, which post-processes each model's point forecasts by adding empirical quantiles of past out-of-sample forecast errors \citep{williams1971simple}. Let $m$ denote the forecasting model, $t$ the forecast origin, and $h$ the forecast horizon. Let $\widehat{Y}_{t,h}^{(m)}$ denote the point forecast of the sea level $Y_{t,h}$, observed $h$ hours after origin $t$. The corresponding out-of-sample forecast error is $e_{t,h}^{(m)}=Y_{t,h}-\widehat{Y}_{t,h}^{(m)}$. At forecast origin $t$, only errors for which the corresponding sea-level observations are available can be used. A rolling window of 60 days ($L=1440$ most recent available forecast errors) is used for each station, model and horizon $C_{t,h}^{(m)}=\{j:t-h-L<j\leq t-h\}$.
The window is updated once a day, so the same set of errors is used for all forecast origins within a given day. Since a complete window is required for every horizon, quantile forecasts are evaluated from 6 March 2023 to the end of the test set. For a quantile level $\tau\in(0,1)$, the quantile forecast is obtained by adding the empirical $\tau$-quantile of the errors in the window to the current point forecast:
\begin{equation}
\widehat{q}_{t,h}^{(m),\tau}
=
\widehat{Y}_{t,h}^{(m)}
+
\widehat{Q}_{\tau}
\left(
\{e_{j,h}^{(m)}:j\in C_{t,h}^{(m)}\}
\right),
\label{eq:empirical_quantile_forecast}
\end{equation}
where $\widehat{Q}_{\tau}$ denotes the empirical $\tau$-quantile, computed according to Definition~7 of \cite{hyndman1996sample}. Thus, quantiles are estimated separately for each station, model, and horizon, using only past out-of-sample forecast errors \citep{lee2014empirical}. Two quantile levels, $\tau\in\{0.90,0.95\}$, are considered at the operationally relevant horizons $h=24$ and $h=48$. An exceedance occurs when the observed sea level is above the upper quantile forecast. The exceedance frequency is defined as
\begin{equation}
\pi_{h}^{(m),\tau}=\frac{1}{|\mathcal{T}_{h}|}\sum_{t\in\mathcal{T}_{h}}I_{t,h}^{(m),\tau},
\end{equation}
where $I_{t,h}^{(m),\tau}=\mathbb{I}\left(Y_{t,h}>\widehat{q}_{t,h}^{(m),\tau}\right)$, and $\mathcal{T}_{h}$ is the set of evaluation origins. For a calibrated quantile forecast, the exceedance frequency should be close to its nominal value, $1-\tau$. Values above (below) $1-\tau$ indicate that the forecast threshold is too low (high), and therefore that the right tail of the sea-level distribution is underestimated (overestimated). The exceedance frequency from its nominal value is $EX^{m,\tau}_h=\pi_{h}^{(m),\tau}-(1-\tau)$. Forecast accuracy is also assessed using the quantile score:
\begin{equation}
{\mathrm{QS}}_{h}^{(m),\tau}=
\frac{1}{T_{h}}\sum_{t\in\mathcal{T}_{h}}
\rho_{\tau}\!\left(Y_{t,h}-\widehat{q}_{t,h}^{(m),\tau}\right),
\label{eq:quantile_score}
\end{equation}
where $\rho_\tau(u)=u\{\tau-\mathbb{I}(u<0)\}$ is the quantile loss. The quantile score is a strictly consistent scoring function for the $\tau$-quantile \citep{gneiting2011quantiles,zaborowski2026}; lower values indicate more accurate upper quantile forecasts. Correct unconditional coverage is tested with the likelihood-ratio test of \citet{Kupiec1995}. Since forecasts are issued every hour, consecutive exceedance indicators overlap and tend to cluster during surge episodes, which violates the independence assumption of the standard test. A HAC-robust version is therefore used, based on the statistic $z=\{\pi_h^{(m),\tau}-(1-\tau)\}/\sqrt{\widehat{V}/T_h}$, where $\widehat{V}$ is the Newey--West long-run variance of the exceedance indicators, computed with a Bartlett kernel and a lag of 167 hours \citep{NeweyWest1987}. Differences in quantile score are assessed with the MCS procedure described above, applied to the quantile loss at each forecast origin.

\section{Forecasting results}
\label{sec:forecasting_results}
\begin{figure}[htbp]
    \centering
    \includegraphics[width=1\textwidth]{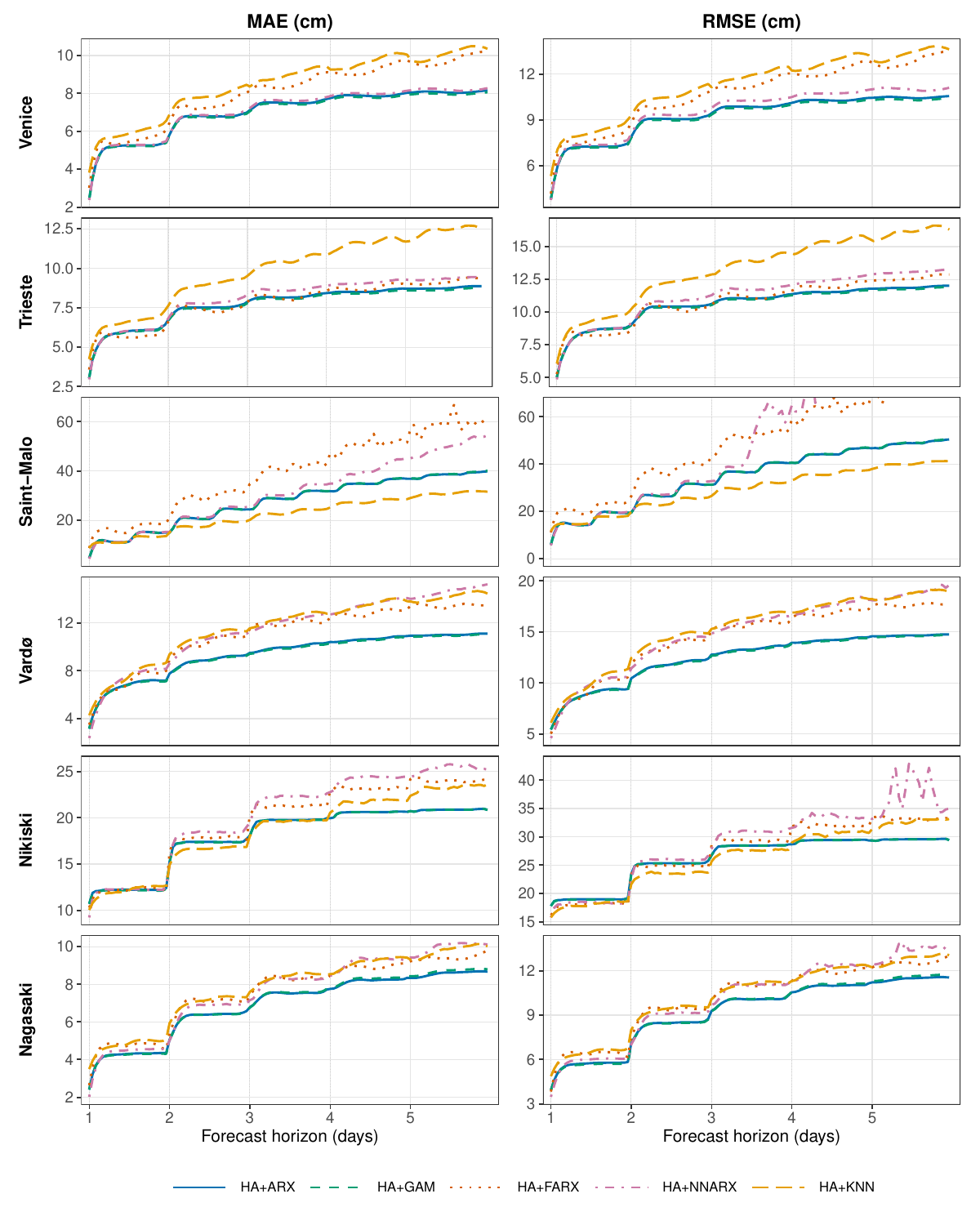}
\caption{Out-of-sample forecast errors for the five hybrid models across forecast horizons of up to five days.}
    \label{fig:mae_rmse_oos}
\end{figure}

\begin{figure}[H]
    \centering
    \includegraphics[width=1\textwidth]{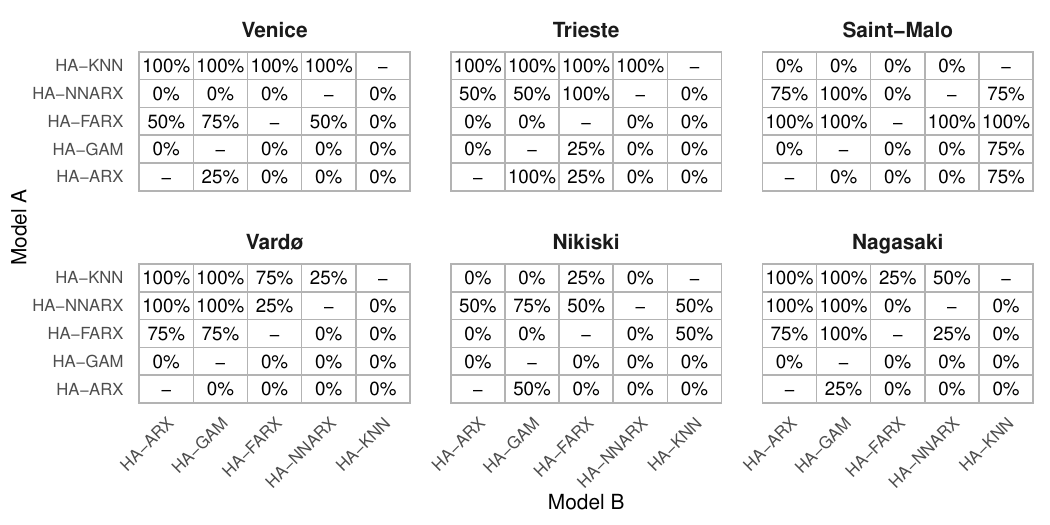}
    \caption{Pairwise Diebold--Mariano test results based on the MAE loss. Each cell reports the percentage at which Model B significantly outperforms Model A (at the 5\% level).}
    \label{fig:dm_heatmap}
\end{figure}

The out-of-sample forecasting errors, in terms of MAE and RMSE (in cm), for the five hybrid models over forecast horizons of up to five days are reported in Figure~\ref{fig:mae_rmse_oos}. As expected, forecast errors increase with the prediction horizon for all specifications, with the most pronounced increase occurring during the first two days. In Venice HA+GAM, followed by HA+ARX provide the lowest MAE and RMSE, while HA+NNARX performs slightly worse. HA+FARX and HA+KNN show higher errors, with the gap becoming more pronounced at longer horizons. In Trieste, HA+ARX and HA+GAM provide the lowest errors, while HA+KNN performs worst. Saint-Malo shows a different ranking, with HA+KNN performing well. For Vard{\o}, HA+ARX and HA+GAM remain the most accurate specifications, whereas Nikiski shows relatively similar performance across models, apart from larger RMSE values for HA+NNARX at long horizons. In Nagasaki, HA+ARX and HA+GAM again provide the lowest errors. 
\noindent
Table~\ref{tab:best_models_12_24_36_48} complements the graphical evidence by reporting the out-of-sample MAE and RMSE of the selected hybrid model at the relevant forecast horizons of 12, 24, 36, and 48 hours. HA+GAM is selected for Venice, Trieste, Vard{\o}, and Nikiski, whereas HA+KNN and HA+ARX are preferred for Saint-Malo and Nagasaki, respectively.
\begin{table}[H]
\centering
\caption{Out-of-sample forecast errors of the selected hybrid models at 12, 24, 36, and 48 hour horizons.}
\label{tab:best_models_12_24_36_48}
\resizebox{\textwidth}{!}{%
\begin{tabular}{lccccccccc}
\toprule
& & \multicolumn{4}{c}{MAE} & \multicolumn{4}{c}{RMSE} \\
\cmidrule(lr){3-6} \cmidrule(lr){7-10}
Station & Model & $h=12$ & $h=24$ & $h=36$ & $h=48$ & $h=12$ & $h=24$ & $h=36$ & $h=48$ \\
\midrule
Venice & HA+GAM & 5.22 & 5.40 & 6.74 & 6.92 & 7.19 & 7.40 & 8.98 & 9.20 \\
Trieste & HA+GAM & 5.94 & 6.36 & 7.43 & 7.66 & 8.58 & 9.09 & 10.35 & 10.55 \\
Saint--Malo & HA+KNN & 11.23 & 13.63 & 17.42 & 19.66 & 14.99 & 18.11 & 22.93 & 25.90 \\
Vard{\o} & HA+GAM & 6.76 & 7.12 & 8.81 & 9.26 & 8.88 & 9.33 & 11.68 & 12.28 \\
Nikiski & HA+GAM & 12.16 & 12.31 & 17.35 & 17.66 & 18.89 & 19.06 & 25.30 & 25.70 \\
Nagasaki & HA+ARX & 4.30 & 4.35 & 6.39 & 6.58 & 5.74 & 5.83 & 8.46 & 8.75 \\
\bottomrule
\end{tabular}%
}
\end{table}
Table~\ref{tab:error_comparison_24} compares the out-of-sample forecast errors of the HA with those of the selected hybrid model at the 24-hour horizon. 
\begin{table}[H]
\centering
\caption{Out-of-sample forecast errors at the 24-hour horizon. HA is harmonic analysis, Hybrid the selected hybrid model, and $\Delta$ the absolute differences between forecast errors.}
\label{tab:error_comparison_24}
\resizebox{\textwidth}{!}{%
\begin{tabular}{lccccccccc}
\toprule
& & \multicolumn{4}{c}{MAE} & \multicolumn{4}{c}{RMSE} \\
\cmidrule(lr){3-6} \cmidrule(lr){7-10}
Station & Model
& HA & Hybrid & $\Delta$ & $\Delta\%$
& HA & Hybrid & $\Delta$ & $\Delta\%$ \\
\midrule
Venice      & HA+GAM & 12.06 &  5.40 &  -6.66 & -55.2 & 16.34 &  7.40 &  -8.94 & -54.7 \\
Trieste     & HA+GAM & 12.88 &  6.36 &  -6.52 & -50.6 & 18.22 &  9.09 &  -9.13 & -50.1 \\
Saint--Malo & HA+KNN & 34.42 & 13.63 & -20.79 & -60.4 & 43.46 & 18.11 & -25.35 & -58.3 \\
Vard{\o}    & HA+GAM & 13.14 &  7.12 &  -6.02 & -45.8 & 17.39 &  9.33 &  -8.06 & -46.3 \\
Nikiski     & HA+GAM & 23.83 & 12.31 & -11.52 & -48.3 & 32.71 & 19.06 & -13.65 & -41.7 \\
Nagasaki    & HA+ARX & 13.93 &  4.35 &  -9.58 & -68.8 & 17.13 &  5.83 & -11.30 & -66.0 \\
\bottomrule
\end{tabular}
}
\end{table}
\noindent
In Venice, HA+GAM reduces the MAE from 12.06 to 5.40 ($-55.2\%$) and the RMSE from 16.34 to 7.40 ($-54.7\%$). In Trieste, the corresponding errors decrease from 12.88 to 6.36 ($-50.6\%$) and from 18.22 to 9.09 ($-50.1\%$). In Saint--Malo, HA+KNN reduces the MAE from 34.42 to 13.63 ($-60.4\%$) and the RMSE from 43.46 to 18.11 ($-58.3\%$). For Vard{\o}, HA+GAM lowers the MAE from 13.14 to 7.12 ($-45.8\%$) and the RMSE from 17.39 to 9.33 ($-46.3\%$). In Nikiski, it reduces the two measures from 23.83 to 12.31 ($-48.3\%$) and from 32.71 to 19.06 ($-41.7\%$), respectively. The largest relative improvement is observed in Nagasaki, where HA+ARX decreases the MAE from 13.93 to 4.35 ($-68.8\%$) and the RMSE from 17.13 to 5.83 ($-66.0\%$). At the 24-hour forecast horizon, the selected hybrid models reduce the MAE and RMSE by an average of 54.9\% and 52.9\%, across the six stations compared with the HA.\\
Figure~\ref{fig:dm_heatmap} reports the results of the pairwise DM tests based on the MAE. Each cell shows the percentage of the four selected forecast horizons 12, 24, 36, and 48 hours at which Model B significantly outperforms Model A after applying the Benjamini--Yekutieli correction for multiple testing. In Venice, HA+ARX, HA+GAM, HA+FARX, and HA+NNARX all outperform HA+KNN at every selected horizon. HA+GAM also outperforms HA+FARX at 75\% of the horizons, while HA+ARX and HA+NNARX do so at 50\%. In Trieste, all four competing models outperform HA+KNN across all horizons, while HA+GAM consistently outperforms HA+ARX and HA+FARX outperforms HA+NNARX. Saint-Malo displays a markedly different ranking: HA+KNN outperforms HA+ARX, HA+GAM, and HA+NNARX at 75\% of the horizons and HA+FARX at all horizons. HA+FARX is also outperformed by each of the other models at every selected horizon. At Vard{\o}, HA+ARX and HA+GAM exhibit the strongest performance, both outperforming HA+KNN and HA+NNARX across all horizons and HA+FARX at 75\% of them. In Nikiski, HA+GAM outperforms HA+NNARX at 75\% of the horizons and HA+ARX at 50\%. In Nagasaki, HA+GAM outperforms HA+FARX, HA+NNARX, and HA+KNN across all four horizons, while HA+ARX also outperforms HA+NNARX and HA+KNN at every horizon and HA+FARX at 75\% of them.

Table~\ref{tab:mcs_mae_h24_h48_results} reports the MCS results based on the MAE loss at the 24- and 48-hour forecast horizons. At 24 hours, the procedure identifies highly selective superior sets: a single model is retained at five of the six stations, while only Venice has a two-model superior set comprising HA+GAM and HA+NNARX. The selected model varies across locations. HA+GAM ranks first in Venice, Vard{\o}, and Nagasaki, HA+FARX in Trieste and Nikiski, and HA+KNN in Saint-Malo. This heterogeneity confirms that no single hybrid specification dominates at every station. At the 48-hour horizon, the superior sets generally become larger, suggesting that differences in predictive ability are more difficult to distinguish as forecast uncertainty increases. The composition of the superior set remains unchanged in Venice and in Saint-Malo. In Trieste, HA+FARX continues to rank first, while HA+GAM and HA+ARX also enter the superior set. Similarly, HA+ARX is retained alongside HA+GAM in Vard{\o} and Nagasaki. Nikiski exhibits the most notable change: HA+KNN ranks first at 48 hours, while HA+FARX and HA+ARX are also retained. Overall, HA+GAM displays the greatest cross-station consistency, ranking first at both horizons in Venice, Vard{\o}, and Nagasaki. Notably, the harmonic model is excluded from every superior set at both horizons, providing further evidence that modelling the non-astronomical component improves forecast accuracy.

\begin{table}[H]
\caption{Model confidence set results based on absolute-error loss at the 24- and 48-hour forecast horizons. Superior sets are constructed for $\alpha=0.15$, corresponding to an 85\% confidence level. MCS p-values are reported in parentheses. The superior set contains models whose predictive performance cannot be statistically distinguished at the 15\% significance level. The last row reports the number of models excluded from each superior set.}
\label{tab:mcs_mae_h24_h48_results}
\centering
\resizebox{0.90\textwidth}{!}{
\begin{tabular}{ccccccc}
\hline
Rank & Venice & Trieste & Saint-Malo & Vard{\o} & Nikiski & Nagasaki \\
\hline
\multicolumn{7}{c}{24-hour forecast horizon} \\
\hline
1
& \makecell{HA+GAM \\ (1.00)}
& \makecell{HA+FARX \\ (1.00)}
& \makecell{HA+KNN \\ (1.00)}
& \makecell{HA+GAM \\ (1.00)}
& \makecell{HA+FARX \\ (1.00)}
& \makecell{HA+GAM \\ (1.00)}
\\
2
& \makecell{HA+NNARX \\ (0.37)}
& -
& -
& -
& -
& -
\\
Excl.
& 4
& 5
& 5
& 5
& 5
& 5
\\
\hline
\multicolumn{7}{c}{48-hour forecast horizon} \\
\hline
1
& \makecell{HA+GAM \\ (1.00)}
& \makecell{HA+FARX \\ (1.00)}
& \makecell{HA+KNN \\ (1.00)}
& \makecell{HA+GAM \\ (1.00)}
& \makecell{HA+KNN \\ (1.00)}
& \makecell{HA+GAM \\ (1.00)}
\\
2
& \makecell{HA+NNARX \\ (0.26)}
& \makecell{HA+GAM \\ (0.86)}
& -
& \makecell{HA+ARX \\ (0.21)}
& \makecell{HA+FARX \\ (0.21)}
& \makecell{HA+ARX \\ (0.28)}
\\
3
& -
& \makecell{HA+ARX \\ (0.27)}
& -
& -
& \makecell{HA+ARX \\ (0.21)}
& -
\\
Excl.
& 4
& 3
& 5
& 4
& 3
& 4
\\
\hline
\end{tabular}
}
\end{table}
Table~\ref{tab:gfs_robustness} reports the out-of-sample forecast errors (cm) for Venice at the 24- and 48-hour horizons with meteorological covariates set to their observed values (Obs.) and to the GFS forecasts. Findings show that GFS forecasts leaves forecast accuracy unchanged. None of the DM tests rejects the null hypothesis of equal predictive accuracy ($p \geq 0.22$), and the ranking of the models is also unchanged. For HA+FARX and HA+KNN at 24 hours the two sets of forecasts coincide, because at this horizon both models rely only on meteorological values already observed at the origin; these models provide operational forecasts up to one day ahead without requiring weather forecasts. At 48 hours they use GFS inputs, and individual forecasts change, but the differences average out. These results indicate that, up to 48 hours, the ex-post design does not favour the specifications that make greater use of
future meteorological information, and that the reported accuracy is representative of what a forecaster could achieve in real time in Venice, for instance when deciding on the activation of MoSE one or two days in advance.

\begin{table}[H]
\centering
\caption{Out-of-sample forecast errors (cm) for Venice at the 24- and 48-hour horizons with meteorological covariates set to their observed values (Obs.) and with GFS forecasts. $\Delta\%$ is the percentage change.}
\label{tab:gfs_robustness}
\small
\begin{tabular}{llcccccccc}
\toprule
 & & \multicolumn{3}{c}{MAE} & \multicolumn{3}{c}{RMSE} & \multicolumn{2}{c}{DM test} \\
\cmidrule(lr){3-5}\cmidrule(lr){6-8}\cmidrule(lr){9-10}
Horizon & Model & Obs. & GFS & $\Delta\%$ & Obs. & GFS & $\Delta\%$ & Stat. & $p$-value \\
\midrule
\multirow{5}{*}{$h=24$}
 & HA+ARX   & 5.43 & 5.45 & $+0.36$ & 7.46 & 7.50 & $+0.57$ & $0.79$  & 0.22 \\
 & HA+GAM   & 5.40 & 5.40 & $+0.13$ & 7.40 & 7.41 & $+0.15$ & $0.26$  & 0.40 \\
 & HA+NNARX & 5.47 & 5.47 & $0.00$  & 7.68 & 7.69 & $+0.12$ & $0.01$  & 0.50 \\
 & HA+KNN   & 6.55 & 6.55 & $0.00$  & 9.02 & 9.02 & $0.00$  & --      & --   \\
 & HA+FARX  & 6.08 & 6.08 & $0.00$  & 8.41 & 8.41 & $0.00$  & --      & --   \\
\midrule
\multirow{5}{*}{$h=48$}
 & HA+ARX   & 6.98 & 6.98 & $-0.04$ & 9.28  & 9.37  & $+0.94$ & $-0.06$ & 0.52 \\
 & HA+GAM   & 6.92 & 6.91 & $-0.26$ & 9.20  & 9.26  & $+0.64$ & $-0.38$ & 0.65 \\
 & HA+NNARX & 7.08 & 7.07 & $-0.14$ & 9.62  & 9.65  & $+0.30$ & $-0.22$ & 0.59 \\
 & HA+KNN   & 8.46 & 8.46 & $0.00$  & 11.35 & 11.35 & $+0.02$ & $-0.02$ & 0.51 \\
 & HA+FARX  & 8.04 & 8.04 & $+0.02$ & 10.81 & 10.81 & $-0.05$ & $0.10$  & 0.46 \\
\bottomrule
\end{tabular}
\end{table}

Tables~\ref{tab:upper_quantile_h24} and \ref{tab:upper_quantile_h48} report the out-of-sample performance of the 90th and 95th percentile forecasts at the 24- and 48-hour horizons.
\begin{table}[H]
\centering
\caption{Accuracy of the upper quantile forecasts by station and quantile level at the 24-hour horizon. Exceedance frequency (EX) from the nominal value ($\%$). Average quantile score (QS) in cm. The EX closest to zero is in bold. $^{*}$: the HAC-robust Kupiec test rejects correct unconditional coverage at the 5\% level. $^{\dagger}$: model included in the 85\% model confidence set based on QS.}
\label{tab:upper_quantile_h24}
\small
\resizebox{\textwidth}{!}{%
\begin{tabular}{llcccccccccccc}
\toprule
& & \multicolumn{2}{c}{HA} & \multicolumn{2}{c}{HA+ARX} & \multicolumn{2}{c}{HA+GAM} & \multicolumn{2}{c}{HA+FARX} & \multicolumn{2}{c}{HA+NNARX} & \multicolumn{2}{c}{HA+KNN} \\
\cmidrule(lr){3-4}\cmidrule(lr){5-6}\cmidrule(lr){7-8}\cmidrule(lr){9-10}\cmidrule(lr){11-12}\cmidrule(lr){13-14}
Station & $\tau$ & EX & QS & EX & QS & EX & QS & EX & QS & EX & QS & EX & QS \\
\midrule
Venice & 0.90 & +2.47 & 2.455 & +2.48 & 1.271$^{\dagger}$ & +2.38 & 1.266$^{\dagger}$ & \textbf{+0.08} & 1.525 & +2.04 & 1.318 & +0.53 & 1.672 \\
 & 0.95 & +2.24 & 1.499 & +1.11 & 0.811$^{\dagger}$ & +1.18 & 0.805$^{\dagger}$ & \textbf{+0.34} & 0.992 & +1.28 & 0.844 & +0.75 & 1.073 \\
\addlinespace
Trieste & 0.90 & +3.60 & 2.993 & +1.59 & 1.645$^{\dagger}$ & +1.47 & 1.646$^{\dagger}$ & +0.90 & 1.661$^{\dagger}$ & +1.79 & 1.695 & \textbf{+0.14} & 2.046 \\
 & 0.95 & +2.96 & 1.879 & +1.06 & 1.127$^{\dagger}$ & +1.10 & 1.131$^{\dagger}$ & +1.06 & 1.119$^{\dagger}$ & +1.07 & 1.138$^{\dagger}$ & \textbf{+0.62} & 1.358 \\
\addlinespace
Saint-Malo & 0.90 & +0.76 & 7.426 & $-$0.14 & 3.472 & $-$0.47 & 3.425$^{\dagger}$ & \textbf{$-$0.12} & 4.200 & $-$0.50 & 3.511 & +1.40 & 3.288$^{\dagger}$ \\
 & 0.95 & +1.23 & 4.268 & +0.31 & 2.094 & +0.35 & 2.055$^{\dagger}$ & +0.20 & 2.507 & \textbf{+0.08} & 2.110$^{\dagger}$ & +1.14 & 2.107$^{\dagger}$ \\
\addlinespace
Vard{\o} & 0.90 & +2.86 & 3.364 & $-$0.36 & 1.556$^{\dagger}$ & $-$0.17 & 1.544$^{\dagger}$ & $-$0.38 & 1.876 & $-$0.44 & 1.702 & \textbf{$-$0.05} & 2.005 \\
 & 0.95 & +2.64 & 2.152 & \textbf{+0.11} & 0.975$^{\dagger}$ & +0.14 & 0.967$^{\dagger}$ & +0.45 & 1.162 & +0.17 & 1.057 & +0.92 & 1.265 \\
\addlinespace
Nikiski & 0.90 & +2.07 & 5.204 & +0.73 & 3.022$^{\dagger}$ & \textbf{+0.48} & 2.994$^{\dagger}$ & +0.80 & 2.876$^{\dagger}$ & +0.55 & 2.890$^{\dagger}$ & +1.30 & 2.947$^{\dagger}$ \\
 & 0.95 & +1.30 & 3.306 & +0.51 & 1.977 & \textbf{+0.44} & 1.957 & +1.26 & 1.784$^{\dagger}$ & +0.46 & 1.817$^{\dagger}$ & +1.18 & 1.882$^{\dagger}$ \\
\addlinespace
Nagasaki & 0.90 & +6.28$^{*}$ & 2.365 & \textbf{$-$0.02} & 0.961 & +0.04 & 0.946$^{\dagger}$ & +0.17 & 1.047 & +0.07 & 1.002 & +0.19 & 1.169 \\
 & 0.95 & +4.05$^{*}$ & 1.446 & +0.08 & 0.610 & \textbf{+0.01} & 0.598$^{\dagger}$ & +0.35 & 0.666 & +0.08 & 0.628 & +0.34 & 0.723 \\
\bottomrule
\end{tabular}}
\end{table}
\noindent
For HA, $EX$ is positive at every station and quantile level, ranging from $+0.76$ to $+6.28$ points at 24 hours and from $+0.83$ to $+6.70$ points at 48 hours. Thus, observed sea levels exceed the HA quantiles more often than expected, indicating that HA underestimates the right tail. The hybrid models generally bring $EX$ closer to zero. For example, at the 90th percentile in Nagasaki, $EX$ falls from $+6.28$ points for HA to $+0.04$ for HA+GAM at 24 hours, and from $+6.70$ to $+0.03$ for HA+NNARX at 48 hours. Negative $EX$ values indicate that the quantile forecast exceeds its nominal coverage. Calibration alone, however, does not determine forecast quality: a threshold set too high may achieve nominal coverage while providing little information. The QS accounts for both exceedances and non-exceedances, weighting their distances from the forecast quantile. Lower QS therefore indicates a more accurate and sharper forecast. Every hybrid model improves on HA in QS at every station, quantile level and horizon. Relative to HA, the best hybrid reduces QS by 40.4--60.0\% at 24 hours and by 26.7--41.9\% at 48 hours. At the 90th percentile in Vard{\o} at 24 hours, HA+GAM has an $EX$ of $-0.17$ points and reduces QS by 54.1\% relative to HA. The model with $EX$ closest to zero also has the lowest QS in only one of the 24 station--quantile--horizon combinations. In most cases, a model with a somewhat larger calibration error produces a sharper forecast and a lower QS. At the 90th percentile in Venice at 24 hours, for example, HA+FARX is best calibrated ($EX=+0.08$), whereas HA+GAM achieves the larger QS reduction (48.4\%, compared with 37.9\% for
HA+FARX). The model with the lowest QS varies across stations and horizons. HA+GAM performs best most often, particularly in Venice, Vard{\o} and Nagasaki, while other specifications lead in Trieste, Saint--Malo and Nikiski. The improvements over HA are generally smaller at 48 hours, consistent with the larger point-forecast errors at the longer horizon.
\begin{table}[H]
\centering
\caption{Accuracy of the upper quantile forecasts by station and quantile level at the 48-hour horizon. Exceedance frequency (EX) from the nominal value ($\%$). Average quantile score (QS) in cm. The EX closest to zero is in bold. $^{*}$: the HAC-robust Kupiec test rejects correct unconditional coverage at the 5\% level. $^{\dagger}$: model included in the 85\% model confidence set based on QS.}
\label{tab:upper_quantile_h48}
\small
\resizebox{\textwidth}{!}{%
\begin{tabular}{llcccccccccccc}
\toprule
& & \multicolumn{2}{c}{HA} & \multicolumn{2}{c}{HA+ARX} & \multicolumn{2}{c}{HA+GAM} & \multicolumn{2}{c}{HA+FARX} & \multicolumn{2}{c}{HA+NNARX} & \multicolumn{2}{c}{HA+KNN} \\
\cmidrule(lr){3-4}\cmidrule(lr){5-6}\cmidrule(lr){7-8}\cmidrule(lr){9-10}\cmidrule(lr){11-12}\cmidrule(lr){13-14}
Station & $\tau$ & EX & QS & EX & QS & EX & QS & EX & QS & EX & QS & EX & QS \\
\midrule
Venice & 0.90 & +2.75 & 2.518 & +3.02 & 1.576 & +3.02 & 1.561$^{\dagger}$ & \textbf{+0.43} & 1.961 & +3.17 & 1.632 & +0.62 & 2.066 \\
 & 0.95 & +2.57 & 1.543 & +2.23 & 0.999 & +1.95 & 0.986$^{\dagger}$ & \textbf{+0.38} & 1.253 & +1.98 & 1.033 & +0.83 & 1.313 \\
\addlinespace
Trieste & 0.90 & +3.74 & 3.082 & +2.04 & 1.927$^{\dagger}$ & +1.65 & 1.927$^{\dagger}$ & +1.42 & 2.004 & +1.75 & 2.072 & \textbf{+0.69} & 2.513 \\
 & 0.95 & +3.43 & 1.949 & +1.51 & 1.267$^{\dagger}$ & +1.30 & 1.269$^{\dagger}$ & +1.20 & 1.292$^{\dagger}$ & +1.62 & 1.347 & \textbf{+0.96} & 1.623 \\
\addlinespace
Saint-Malo & 0.90 & +0.83 & 7.472 & $-$0.75 & 5.617 & $-$0.79 & 5.569 & \textbf{$-$0.38} & 7.089 & $-$0.78 & 5.834 & +0.77 & 4.603$^{\dagger}$ \\
 & 0.95 & +1.30 & 4.298 & +0.21 & 3.356 & \textbf{+0.06} & 3.333 & +0.18 & 4.193 & $-$0.14 & 3.496 & +0.76 & 2.884$^{\dagger}$ \\
\addlinespace
Vard{\o} & 0.90 & +3.06 & 3.421 & \textbf{+0.04} & 2.023$^{\dagger}$ & +0.36 & 2.008$^{\dagger}$ & +0.19 & 2.639 & $-$0.34 & 2.378 & +0.46 & 2.736 \\
 & 0.95 & +3.12 & 2.213 & +0.46 & 1.295$^{\dagger}$ & +0.54 & 1.286$^{\dagger}$ & +0.73 & 1.651 & \textbf{+0.10} & 1.501 & +1.42 & 1.760 \\
\addlinespace
Nikiski & 0.90 & +2.12 & 5.226 & \textbf{+0.31} & 4.349 & +0.42 & 4.334 & +1.25 & 4.045 & +0.42 & 4.359 & +1.37 & 3.833$^{\dagger}$ \\
 & 0.95 & +1.34 & 3.323 & \textbf{+0.31} & 2.826 & +0.34 & 2.813 & +1.28 & 2.501$^{\dagger}$ & +0.48 & 2.755 & +1.23 & 2.418$^{\dagger}$ \\
\addlinespace
Nagasaki & 0.90 & +6.70$^{*}$ & 2.408 & +0.42 & 1.447$^{\dagger}$ & +0.41 & 1.437$^{\dagger}$ & +0.14 & 1.613 & \textbf{+0.03} & 1.530 & $-$0.17 & 1.593 \\
 & 0.95 & +4.38$^{*}$ & 1.478 & +0.20 & 0.905$^{\dagger}$ & +0.13 & 0.893$^{\dagger}$ & +0.52 & 1.010 & +0.15 & 0.952 & \textbf{+0.08} & 0.983 \\
\bottomrule
\end{tabular}}
\end{table}
\noindent
The HAC-robust Kupiec test rejects correct coverage at the 5\% level only for HA at Nagasaki, at both quantile levels and horizons. For the hybrid models, the deviations of the exceedance frequency from its nominal value are never statistically significant. The MCS based on the QS never retains HA, confirming that the hybrid models significantly improve the upper quantile forecasts. HA+GAM belongs to the superior set in 11 of the 12 station--quantile combinations at 24 hours and in 8 at 48 hours, followed by HA+ARX (7 and 6). At 48 hours, HA+KNN is the only model retained at Saint-Malo and is retained, together with HA+FARX, at Nikiski, consistently with the MCS results for the point forecasts.

\section{The economic value of tide forecasts for decision making in Venice}
\label{sec:mose_cost_loss}

To illustrate the economic relevance of accurate sea-level forecasts, the case of Venice is considered.  The operational relevance of sea-level forecasting has increased with the introduction of MoSE, a system of mobile flood-barriers installed at the three lagoon inlets to protect the city from high-water events. When a closure is required, the gates are raised to form a temporary barrier between the Adriatic Sea and the Venice Lagoon, thereby limiting the propagation of high sea levels into the lagoon. The system was first activated on 3 October 2020. Since barrier closures must be decided in advance, sea-level forecasts are directly linked to an operational decision-making issue. Each activation entails energy and personnel costs, estimated at approximately \texteuro200,000 per closure \citep{biondi2026}. MoSE was activated 28 times during 2022--2023, with an estimated total operational cost of about \texteuro3.2 million. Forecast errors therefore have asymmetric economic consequences: overestimation may trigger unnecessary closures and additional operating costs, whereas underestimation may leave the city exposed to potentially much larger flood damages. For example, the November 2019 high-water event caused \texteuro93 million in losses and \texteuro82 million in emergency response and structural damage costs \citep{giupponi2024}.

The existing out-of-sample forecasts are evaluated within a cost--loss framework at the 24- and 48-hour horizons. Let $H\in\{24,48\}$ denote the length of the decision window. For each model $m$ and forecast origin $i$, the decision score is defined as the maximum predicted sea level within the subsequent $H$ hours,
\begin{equation}\label{eq:decision_score}
S_{i,H}^{(m)}
=
\max_{1\leq h\leq H}
\widehat{Y}_{i,h}^{(m)}.
\end{equation}
The analysis evaluates whether a model anticipates a potentially critical sea level at any point within the relevant operational window, rather than considering only the point forecast at its endpoint. Given a forecast decision threshold $c$, an activation is implied whenever
\begin{equation}
\widehat{D}_{i,H}^{(m)}(c)
=
\mathbb{I}
\left(
S_{i,H}^{(m)}\geq c
\right).
\end{equation}
Let $C$ denote the cost of a MoSE activation. Based on available estimates of its operating costs, $C$ is set to $\euro 200{,}000$ per activation \citep{giupponi2024,biondi2026}. In contrast, the economic loss is allowed to vary with the severity of the observed high-water event. For each decision window, the maximum observed sea level is defined as
\begin{equation}
Y_{i,H}^{\max}
=
\max_{1\leq h\leq H}
Y_{i,h}.
\end{equation}
and the associated flood damage is
\begin{equation}
L_{i,H}
=
g\left(Y_{i,H}^{\max}\right),
\end{equation}
where $g(\cdot)$ is the estimated flood damage proposed by \cite{caporin2016} and reported in Table~\ref{tab:damage_function}. Accordingly, the economic loss associated with model $m$, decision threshold $c$, and forecast window $H$ is defined as
\begin{equation}\label{eq:loss_function}
EC_{m,H}(c)
=
C\,N_{\mathrm{act},m,H}(c)
+
\sum_{i\in\mathcal{M}_{m,H}(c)}
L_{i,H},
\end{equation}
where $N_{\mathrm{act},m,H}(c)$ is the number of model-implied activations and $\mathcal{M}_{m,H}(c)$ denotes the set of historical MoSE activation windows not identified by model $m$ \citep{comunevenezia2026}. Losses are reported as the mean decision loss per daily decision window, $\overline{EC}_{m,H}(c)=EC_{m,H}(c)/n$, where $n=296$ is the number of daily windows in the test set. Since no activation is needed on most days, $\overline{EC}$ is well below the cost $C$ of a single activation; multiplying it by $n$ gives the total loss over the sample.
\begin{table}[H]
\centering
\caption{Estimated flood damage used in the cost--loss analysis.}
\label{tab:damage_function}
\begin{tabular}{lclc}
\toprule
Sea level (cm) & Estimated damage & Sea level (cm) & Estimated damage \\
               & (\euro\ million) &                & (\euro\ million) \\
\midrule
$\leq 80$ & 0.00   & 131--140 & 177.12 \\
81--90     & 0.56   & 141--150 & 189.16 \\
91--100    & 6.97   & 151--160 & 194.94 \\
101--110   & 23.03  & 161--170 & 195.85 \\
111--120   & 69.08  & 171--180 & 196.07 \\
121--130   & 135.01 & $>180$   & 196.33 \\
\bottomrule
\end{tabular}
\end{table}

\begin{figure}[H]
    \centering
    \includegraphics[width=1\textwidth]{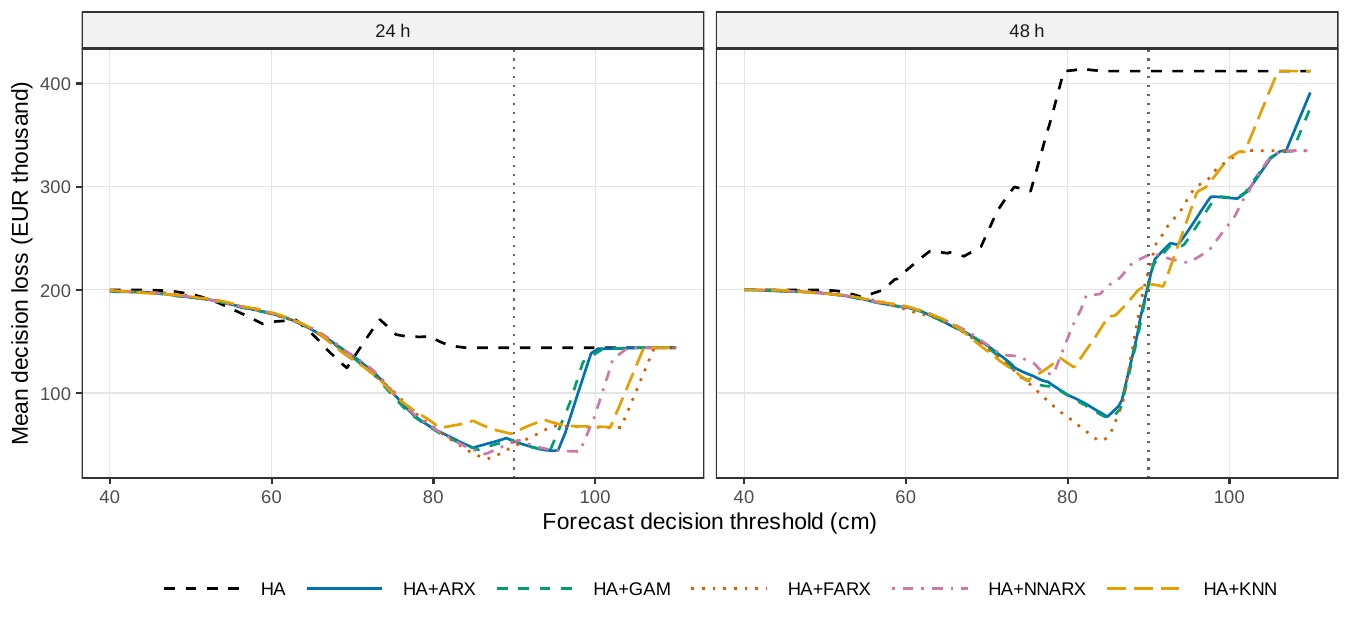}
    \caption{Mean decision loss per daily decision window (\euro thousand) as a function of the forecast decision threshold, for 24- and 48-hour decision windows, 7 March--27 December 2023. Hybrid models use operational GFS weather forecasts. The vertical dotted line indicates the 90~cm high-water reference level.}
    \label{fig:mose_cost_loss}
\end{figure}

\noindent
Each observed MoSE activation is assumed to have been necessary, as the counterfactual sea level that would have occurred in the absence of the flood-barrier system is not available. Missed events are not assigned the same economic consequence: failing to identify a moderate high-water episode produces a substantially smaller loss than missing an extreme event. Since the damage component depends on the maximum observed sea level, the economic value of a given event is common across competing forecasting models. For the hybrid models, the forecasts used in this section are those obtained with the operational GFS weather forecasts available at each forecast origin, so that the cost--loss analysis reflects the information available to the decision maker in real time. Accordingly, their quantile forecasts in Eq.~\eqref{eq:empirical_quantile_forecast} are obtained from the past errors of these forecasts. HA does not use meteorological inputs, and HA+FARX and HA+KNN do not require weather forecasts at the 24-hour horizon. Figure~\ref{fig:mose_cost_loss} reports the mean decision loss as a function of the forecast threshold for the 24- and 48-hour decision windows. At low thresholds, the models imply frequent precautionary activations and operating costs dominate the loss. As the threshold increases, the number of activations declines, but the risk of missing economically damaging high-water events increases, so that at high thresholds the loss is dominated by the damages from missed events. The curves therefore capture the trade-off between precautionary operating costs and severity-dependent damages. At the 24-hour horizon, the hybrid models produce very similar cost--loss profiles and attain their minimum losses at thresholds between approximately 87 and 91~cm, with mean losses between \euro 33 and 56 thousand per decision window. HA remains close to the hybrid specifications up to about 70~cm but performs worse above this level, and its minimum loss, \euro 111 thousand at 71~cm, is substantially higher. At the 48-hour horizon, the differences become more pronounced. The HA curve departs from those of the hybrid models already at about 60~cm, indicating that the astronomical component alone provides limited support for decisions over the longer window, and its minimum loss rises to \euro 186 thousand. Among the hybrid specifications, HA+FARX attains the lowest minimum loss, \euro 45 thousand at a threshold of approximately 86~cm, followed by HA+GAM and HA+ARX, with \euro 68 and 70 thousand at about 87~cm. HA+KNN and HA+NNARX display higher minimum losses, of \euro 99 and 107 thousand, at lower thresholds of 77--79~cm. Incorporating the non-astronomical component therefore substantially improves the economic value of the forecasts, especially at the 48-hour horizon. For most hybrid models, the loss-minimising thresholds are close to the 90~cm high-water reference level \citep{ComuneVeneziaConst}.

These results rely on point forecasts, which ignore the uncertainty around the predicted sea level, although an activation is needed whenever a critical level is likely to be reached. Since the decision concerns the right tail of the sea-level distribution, the same cost--loss framework can be applied to probabilistic forecasts by replacing the decision score in Eq. \eqref{eq:decision_score} with the maximum upper quantile forecast within the decision window,
\begin{equation}
S_{i,H}^{(m),\tau}=\max_{1\le h\le H}\widehat{q}_{i,h}^{(m),\tau},
\label{eq:decision_score_tau}
\end{equation}
where $\widehat{q}_{i,h}^{(m),\tau}$ is the quantile forecast in Eq. \eqref{eq:empirical_quantile_forecast}. An activation is then implied whenever $S_{i,H}^{(m),\tau}\ge c$. Whereas the point forecast rule only requires a threshold $c$, the probabilistic rule also requires a quantile level $\tau$, which determines the margin of protection above the central forecast. As for the point forecasts, $\tau\in\{0.50,0.51,\ldots,0.99\}$ and $c$ are set to the values that minimise the loss in Eq. \eqref{eq:loss_function}. Since the quantile forecasts require a complete window of past errors, both rules are compared over the 296 daily decision windows issued between 7 March and 27 December 2023.
Table~\ref{tab:decision_two_cases_gfs} compares decision rules based on point and quantile forecasts within the MoSE cost--loss framework. For the quantile-based rule, $\tau$ is the quantile level used to construct the decision score in Eq.~\eqref{eq:decision_score_tau}, and $c$ is the activation threshold applied to it.
\begin{table}[H]
\centering
\caption{Cost--loss comparison of decision rules based on point and quantile forecasts obtained with operational GFS weather forecasts, 7 March--27 December 2023 ($n=296$ days). For the point-forecast rule, $c$ (cm) is the threshold that minimises the mean decision loss. For the quantile-forecast rule, $\tau$ and $c$ (cm) jointly minimise it. $\overline{EC}$ is the mean decision loss per daily window, in thousands of euros.}
\label{tab:decision_two_cases_gfs}
\small
\begin{tabular}{lccccc}
\toprule
& \multicolumn{2}{c}{Point forecast} & \multicolumn{3}{c}{Quantile forecast} \\
\cmidrule(lr){2-3}\cmidrule(lr){4-6}
Model & $c$ (cm) & $\overline{EC}$ & $\tau$ & $c$ (cm) & $\overline{EC}$ \\
\midrule
\multicolumn{6}{l}{\textit{$h=24$}} \\
HA        & 71.36 & 111 & 0.97 & 110.10 & 43 \\
HA+ARX    & 86.81 & 40 & 0.95 & 107.31 & 26 \\
HA+GAM    & 86.97 & 39 & 0.92 & 103.71 & 26 \\
HA+FARX   & 86.60 & 33 & 0.80 & 97.05 & 26 \\
HA+NNARX  & 88.33 & 36 & 0.89 & 103.37 & 25 \\
HA+KNN    & 91.49 & 56 & 0.98 & 110.04 & 36 \\
\addlinespace
\multicolumn{6}{l}{\textit{$h=48$}} \\
HA        & 56.29 & 186 & 0.99 & 115.93 & 80 \\
HA+ARX    & 86.81 & 70 & 0.99 & 108.98 & 47 \\
HA+GAM    & 86.97 & 68 & 0.99 & 108.68 & 47 \\
HA+FARX   & 85.62 & 45 & 0.52 & 86.70 & 44 \\
HA+NNARX  & 78.52 & 107 & 0.98 & 104.00 & 53 \\
HA+KNN    & 76.87 & 99 & 0.98 & 104.81 & 91 \\
\bottomrule
\end{tabular}
\end{table}
\noindent
At the 24-hour horizon, the quantile-based rule yields a lower mean loss than the point-forecast rule for every model. For HA, the selected rule uses the 0.97 quantile forecast and a threshold of 110.10~cm. It implies 53 activations and misses 6 MoSE windows, for a mean loss of \euro 43 thousand per window (\euro 12.8 million in total), against \euro 111 thousand under the point-forecast rule. The lowest mean loss, \euro 25 thousand, is achieved by HA+NNARX, closely followed by HA+ARX, HA+GAM and HA+FARX with \euro 26 thousand. At the 48-hour horizon, the quantile-based rule again reduces the mean loss for all six models, although only marginally for HA+FARX. The largest reductions are obtained for HA and HA+NNARX, whose mean losses fall from \euro 186 to 80 thousand and from \euro 107 to 53 thousand, respectively. The lowest mean loss, \euro 44 thousand, is achieved by HA+FARX with the quantile-based rule, against \euro 45 thousand under the point-forecast rule. The selected quantile levels are mostly high, particularly at 48 hours, and the corresponding thresholds generally exceed 100~cm, the main exception being HA+FARX at 48 hours ($\tau=0.52$, $c=86.70$~cm). Overall, accounting for forecast uncertainty improves the economic performance of the activation rule, although the size of the gain depends on the model and the horizon.

\section{Conclusions}
\label{sec:conclusions}

This study develops and compares a set of hybrid models for hourly sea-level forecasting across six tide-gauge stations characterized by different coastal and tidal regimes. The proposed framework decomposes observed sea level into an astronomical component, estimated through HA, and a non-astronomical component, modelled using ARX, GAM, NNARX, KNN, and FARX specifications. The empirical results show substantial heterogeneity in model performance across stations and forecast horizons. Nevertheless, incorporating the non-astronomical component consistently improves forecast accuracy relative to HA alone. At the 24-hour horizon, the selected hybrid models reduce the MAE and RMSE by an average of 54.9\% and 52.9\%, respectively. HA+GAM provides the lowest forecast errors in Venice, Trieste, Vard{\o}, and Nikiski, whereas HA+KNN performs best at Saint--Malo and HA+ARX at Nagasaki. The DM test and the MCS procedure confirm that no single hybrid specification dominates uniformly across locations and horizons. An additional forecasting experiment for Venice shows that replacing observed meteorological covariates with the operational GFS forecasts available at each origin has a negligible effect on forecast accuracy at 24 and 48 hours and leaves the ranking of the hybrid models unchanged. These findings highlight the importance of adapting the model specification to local tidal and meteorological conditions. The proposed KNN approach provides a complementary non-parametric specification by selecting both non-astronomical dynamics and meteorological similarity.

The Venice case study further demonstrates the operational and economic relevance of accurate sea-level forecasts. The cost--loss framework translates out-of-sample forecasts into decision losses by combining the cost of precautionary MoSE activations with severity-dependent flood damage from missed high-water events. The hybrid models produce lower decision losses than HA, particularly over the 48-hour decision window. For the hybrid models, the lowest mean decision loss is achieved at thresholds of approximately 87--91~cm for the 24-hour horizon and 77--87~cm for the 48-hour horizon, close to the 90~cm high-water reference level. Activation rules based on upper quantile forecasts reduce the mean decision loss further for all models at both horizons. These findings depend on the underlying cost estimates, as the decision problem involves balancing the costs and benefits to activate the system. Further research could provide a more detailed assessment of both activation costs and the losses associated with failure to activate the system \citep{Faranda2023,michielotto2026,biondi2026}. In addition, $\tau$ and $c$ are selected ex post on a limited number of activation windows, and the probabilistic rule requires one additional parameter, so the reported savings may be optimistic. A preliminary check, in which $\tau$ and $c$ are selected on the remaining calendar months, indicates that the probabilistic rule still yields a lower loss in 9 of the 12 cases, one of which with a negligible difference. Future research could therefore select the decision parameters in an online forecasting framework, using past information and a longer evaluation period with a larger number of activation events.\\
In this work, quantile forecasts are obtained by post-processing the point forecasts of each model through historical simulation of past out-of-sample errors. Future research could instead consider models that directly forecast the predictive distribution of sea level, such as distributional regression or quantile regression models, so that predictive uncertainty is modelled jointly with the non-astronomical component \citep{kavousi2016,jalali2022,zhuge2024}. At the same time, the post-processing approach could be expanded by considering other methods like GARCH and quantile regression models \citep{zaborowski2026}. Given the heterogeneity in model rankings, future extensions could additionally consider adaptive forecast combinations with weights that vary across stations, forecast horizons, and hydrodynamic conditions \citep{TIMMERMANN2006,WANG2023}. Furthermore, the comparison could be extended by including hydrodynamic models and more complex neural-networks, such as LSTM networks \citep{ishida2020,umgiesser2021,almaliki2025}. To conclude, the real-time evaluation based on operational weather
forecasts could be extended to all stations and to horizons beyond 48 hours, to assess how the uncertainty of weather forecasts propagates into the predictive distribution of sea level.

\bibliographystyle{apalike}
\bibliography{References.bib}
\noindent
\textbf{Conflict of interest disclosure} The authors declare no conflicts of interest.\\
\noindent
\textbf{Funding statement} This work received no external funding.\\
\noindent
\textbf{Data availability statement} Hourly sea-level data are available from the University of Hawaii Sea Level Center (Trieste, Vard{\o}, Nikiski, Nagasaki), Shom (Saint-Malo) and the Municipality of Venice (Venice). Meteorological observations and forecasts are available from NOAA through the R package climate, and GFS, respectively. The code used in the analysis is available from the corresponding author upon request.\\
\noindent
\textbf{Acknowledgments} 
The authors gratefully acknowledge the Municipality of Venice and the Centro Previsioni e Segnalazioni Maree for the data and materials on tides.




\end{document}